\documentclass{jfm}

\usepackage{graphicx}
\usepackage{subcaption}
\usepackage{newtxtext}
\usepackage{newtxmath}
\usepackage{natbib}
\usepackage{hyperref}
\hypersetup{
    colorlinks = true,
    urlcolor   = blue,
    citecolor  = blue
}

\newcommand{\RomanNumeralCaps}[1]
\linenumbers

\title{Predictability of weakly turbulent systems from spatially sparse observations using data assimilation and machine learning}

\author{Vikrant Gupta\aff{1, 2},
Yuanqing Chen\aff{1, 2}
\and
Minping Wan\aff{1, 2, 3}
  \corresp{\email{wanmp@sustech.edu.cn}}}

\affiliation{\aff{1}Guangdong Provincial Key Laboratory of Turbulence Research and Applications, Department of Mechanics and Aerospace Engineering, Southern University of Science and Technology, Shenzhen 518055, PR China,
\aff{2}Guangdong-Hong Kong-Macao Joint Laboratory for Data-Driven Fluid Mechanics and Engineering Applications, Southern University of Science and Technology, Shenzhen, 518055, PR China,
\aff{3}Jiaxing Research Institute, Southern University of Science and Technology, Jiaxing, 314031, PR China}

\begin{document}
\maketitle

\begin{abstract}
We apply two data assimilation (DA) methods, a smoother and a filter, and a model-free machine learning (ML) shallow network to forecast two weakly turbulent systems. We analyse the effect of the spatial sparsity of observations on accuracy of the predictions obtained from these data-driven methods. Based on the results, we divide the spatial sparsity levels in three zones. First is the good-predictions zone in which both DA and ML methods work. We find that in the good-predictions zone the observations remain dense enough to accurately capture the fractal manifold of the system's dynamics, which is measured using the correlation dimension. The accuracy of the DA methods in this zone remains almost as good as for full-resolution observations. Second is the reasonable-predictions zone in which the DA methods still work but at reduced prediction accuracy. Third is the bad-predictions zone in which even the DA methods fail.
We find that the sparsity level up to which the DA methods work is almost the same up to which chaos synchronisation of these systems can be achieved. The main implications of these results are that they (i) firmly establish the spatial resolution up to which the data-driven methods can be utilised, (ii) provide measures to determine if adding more sensors will improve the predictions, and (iii) quantify the advantage (in terms of the required measurement resolution) of using the governing equations within data-driven methods. We also discuss the applicability of these results to fully developed turbulence. 
\end{abstract}

\begin{keywords}
Authors should not enter keywords on the manuscript.
\end{keywords}


\section{Introduction}
\label{Int}


Prediction of turbulent flows from limited observations is of interest in several geophysical and engineering systems. Reduced-order methods based on rapid-distortion theory~\citep{Mann1994}, input-output coherence~\citep{Adrian1988, Bonnet1998}, and low-rank approximations~\citep{Holmes2012, Illingworth2018} are used in various practical applications.
Recently, data assimilation (DA)~\citep{Colburn2011, Li2019, Wang2021, Bauwer2021, He2024} and machine learning (ML)~\citep{Fukami2019, Liu2020, Kim2021, Vlachas2022} are gaining popularity for turbulence prediction.
Both DA and ML can use all available observations to find an optimal estimate of the system state and can be interpreted in Bayesian framework~\citep{Bonavita2021}.
With increasing computational resources, these methods can potentially provide highly accurate predictions even in complex flow systems.
However, DA and ML may have conditions on the resolution of observations to have any advantage over the reduced-order methods~\citep{Suzuki2017}. The present study examines for weakly turbulent systems how the prediction accuracy of some popular DA and ML methods varies with the spatial resolution of the observations.

\subsection{Data assimilation for turbulence prediction}

Initially developed for numerical weather prediction, DA encompasses mathematical methods designed to predict chaotic systems~\citep{Bouttier1999}. Such systems exhibit sensitive dependence on initial conditions. Consequently, neither regression-based methods relying solely on measurements~\citep{Nelson1998} nor simulation of model equations from roughly estimated initial conditions are suitable for predicting chaotic systems. In DA, measurements and equations are combined such that they constraint each other to keep the predictions faithful to the ground truth dynamics.
Two of the most popular DA methods are 4D-Var (four-dimensional variational methods)~\citep{Schlatter2000} and EnKF (ensemble Kalman filter methods)~\citep{Evensen2003}, which also happen to be the most advanced DA methods applied for predicting three-dimensional turbulent flows to date. In 4D-Var, 4D stands for the fact that all the observations in space and time (maximum four-dimensional) are assimilated together. 4D-Var, therefore, produces maximum a posteriori estimation based on the present and future observations, i.e. it is a smoother~\citep{Bouttier1999}. EnKF is inspired by the Kalman filter. In EnKF, Monte Carlo approach is used for determining the uncertainty in the state estimation, which is then used to correct the predictions. EnKF, therefore, produces a minimum variance estimation from the past observations, i.e. it is a filter. Turbulence literature sometimes suggests 4D-Var to be superior than EnKF~\citep{Wang2021}. However, both methods have their strengths and limitations~\citep{Kalnay2007, Gustafsson2007}, and can even be combined together to get superior results~\citep{Carrassi2018}.


In the past two decades, there are several applications of 4D-Var and EnKF for turbulence prediction~\citep{Chevalier2006, Heitz2010, Colburn2011, Gronskis2013, Hayase2015, Kato2015, Suzuki2017, Li2019, Chandramouli2020, Bauwer2021, Wang2021, Du2023, He2024}. On the one hand, it is clear that the measurement resolution required for accurate reconstruction can be far coarse than that is required for numerical simulations. For example, \citet{Yoshida2005} showed that two Kolmogorov flows can be synchronised (to machine accuracy) by continuously substituting only large wavenumber fluctuations from the master system in the slave system. Such cut-off wavenumber is given by $k_s \approx 0.2\eta^{-1}$, where $\eta$ is the Kolmogorov length scale. For similar flows but with different large-scale forcing, \citet{Lalescu2013} determined the cut-off wavenumber to be $k_s \approx 0.15\eta^{-1}$. \citet{Li2024} also found the cut-off wavenumber to vary in this range, i.e. $k_s\approx 0.15-0.20\eta^{-1}$, for a number of rotating turbulent flows in periodic boxes. On the other hand, the measurement resolution cannot be too coarse otherwise DA methods may not show any advantage over reduced-order methods. For example, \citet{Suzuki2017} showed that when estimations in a turbulent channel flow are obtained from wall measurements alone, DA methods do not produce superior results as compared to correlation-based linear stochastic estimation. Such loss in prediction accuracy when reconstruction is attempted from very coarse measurements is also reported in \citet{Li2019} and \citet{Wang2021}.

\citet{Li2019} performed a systematic analysis on the ability of 4D-Var to reconstruct small-scale flow fluctuations from coarse resolution in three-dimensional Kolmogorov flows. Although they did not find a conclusive evidence for a cut-off wavenumber up to which measurements should be available, they found that the small-scale reconstruction is successful when the measurements up to wavenumber of the order $k_s \approx 0.2\eta^{-1}$ are available. \citet{Wang2021} performed a similar study for turbulent channel flows, which are anisotropic and inhomogeneous and are thus significantly more complex. They also found that the Kolmogorov length-scale based criterion (i.e. $k_s \approx 0.2\eta^{-1}$) is indicative of the required measurement resolution. However, they concluded that the Taylor micro-scale relates better with the required resolution for the flow predictability. In their follow-up study on chaos synchronization, \citet{Wang2022} concluded that Taylor micro-scale accounts for the inhomogeneity and anisotropy and thus can give more general conditions for predictability. \citet{DiLeoni2020} used a simple data assimilation technique, nudging, for predicting homogeneous isotropic flow, Rayleigh-B\'{e}nard convection and magnetohydrodynamic flow. They found that the prediction accuracy is influenced by the presence of large-scale coherent structures in the flow as well as by the quality of observations provided. For example, if the observations were collected at fixed spatial locations (Eulerian DA) or were collected by passively moving probes (Lagrangian DA).

\subsection{Machine learning for turbulence prediction}

ML encompasses data-driven methods that do not explicitly need instructions. Their main advantage, therefore, is that they do not necessarily need the model equations. The extensive use of ML methods, particularly neural networks, is recent but the underlying concepts of ML are similar to DA~\citep{Bonavita2021}. In DA, the observations and model equations are used to obtain the system state. In ML, the observations and system state are used to obtain the network model (i.e. weights and biases in the training phase). The difference between the two becomes blurry as DA implements data-driven model correction and ML incorporates the model equations. Most notable among ML that incorporates the model equations is physics-informed neural network (PINN) \citep{Raissi2019, Raissi2020}. PINN is remarkably easy to implement and produces comparable results to 4D-Var even for three-dimensional turbulent flows~\citep{Du2023}. However, a reasonable hypothesis is that if DA and ML are provided with the same model equations, then DA should work better~\citep{Bonavita2021}. This is confirmed by \citet{Du2023}, who notes that beyond the observation horizon the accuracy of PINN deteriorates faster than that from 4D-Var because the model equations in PINN are only satisfied in the L2-sense.

In this work, we limit ourselves to model-free ML so as to maintain a clear distinction between DA and ML. A number of ML methods for turbulence super-resolution~\citep{Fukami2019, Liu2020, Fukami2021, Kim2021, Li2024} and turbulence prediction~\citep{Wan2017, Li2020FNO, Vlachas2022, Racca2023, Li2023} are recently developed. The main focus of these studies has been on innovations in the neural network architecture to achieve physically realistic and accurate predictions. \citet{Fukami2019} combined the use of convolutional neural networks (CNN) and multi-scale layers to capture the multi-scale nature of turbulent flows. \citet{Kim2021} also used CNN but they further added a generative adversarial network to enable unsupervised learning. Their network showed superior performance in its ability to produce physically consistent predictions. To enhance the computational efficiency, \citet{Li2020FNO} used neural operator in the Fourier space while \citet{Vlachas2022} and \citet{Racca2023} employed autoencoders. The latter studies use recurrent neural networks (RNN), which can be easily adapted for predicting the future dynamics of large chaotic systems~\citep{Pathak2018}. Although it is understood from these studies that small-scale reconstruction and prediction become harder as spatio-temporal resolution gets coarser, they do not attempt a systematic analysis of the required resolution. This is mainly because it is difficult to isolate if the prediction errors arise from the lack of resolution or from unsuitability of the network.

\subsection{Contributions of the present work}

There seems to be a close relation between the coarsest resolutions required for chaos synchronisation and for small-scale reconstruction using DA methods. However, to our knowledge, such a relationship has not been established conclusively to date. Our first objective is to find whether such a relationship exists or if DA and ML methods can give skilled predictions beyond the resolution required for chaos synchronisation.
The rate at which chaos synchronisation is achieved gets slower with decreasing resolution. However, the effect of resolution on the prediction accuracy of DA and ML methods is not clear. Our second objective is to analyse the variation in prediction accuracy with changing spatial resolution.
The condition for chaos synchronisation in previous studies is mostly expressed in terms of Kolmogorov length-scale. This is understandable because Kolmogorov length-scale gives an estimate of the smallest length scale in turbulent flows. However, this criterion is difficult to generalise, as found in~\citet{Wang2021}, and can be difficult to calculate.
Our third objective, therefore, is to obtain quantities that can be measured from available observations, such as those from information theory~\citep{Boffetta2002, Lozano2022}, to explain the system's predictability.

Here comes our first challenge - which system should we study? It is tempting to directly dive into three-dimensional turbulent flows (at high Reynolds numbers if possible). However, there are two main problems with such an undertaking. Firstly, these flows are computationally expensive, which makes application of predictive methods as well as further analysis of the prediction results challenging. Secondly, and more importantly, the intermittency (non-Gaussianity) and multi-scale dynamics inherent to three-dimensional turbulence can be beyond the scope of current state-of-the-art DA methods~\citep{Yano2018}. This makes the reason for failure inconclusive, i.e. we do not know whether the failure is because of lack of resolution or lack of skill of the DA method. Drastic loss in performance is also reported when ML methods successful in predicting chaotic systems and two-dimensional turbulent flows are applied to three-dimensional turbulent flows~\citep{Fukami2021}.

In this work, we follow the approach of~\citet{Holmes2012} and study a tractable example in the form of Kuramoto--Sivashinsky (KS) system. This is a spatially extended system, which exhibits interesting dynamics relevant to several flow systems. \citet{Cvitanovic2010} further gives convincing argument for studying KS system over fully developed turbulence. Most engineering and geophysical flows are often dominated by coherent structures and are thus amenable to low-dimensional representation. We also choose to study the complex Ginzburg--Landau (CGL) systems in defect chaos regime (as opposed to phase turbulence regime observed in the KS system) to appropriately generalise the results.
The predictions are obtained using two DA methods, 4D-Var and EnKF, and one ML method, reservoir-computing-based recurrent neural network (RC-RNN)~\citep{Pathak2018, Gupta2023, Racca2023}. These methods are popular as well as powerful enough to predict weakly turbulent systems.

\subsection{Outline}

In Section \ref{Sys}, we present the model equations, numerical methods, and briefly describe the dynamics observed for the KS and CGL systems.
In Section \ref{Meth}, we explain the three methods, 4D-Var, EnKF and RC-RNN, particularly highlighting their key differences and similarities. In Section \ref{Res}, we present the variations in prediction accuracy with spatial resolution of the measurements. In Section \ref{CS}, we obtain the spatial resolution condition for chaos synchronisation and relate that with the DA prediction results. In Section \ref{SD}, we introduce the measures of the system's dynamics in order to explain the effect of spatial sparsity of observations on the prediction accuracy achieved by the DA and ML methods. 
In Section \ref{Con}, we present the main conclusions.

\section{Systems}\label{Sys}

The KS and CGL systems model a variety of flow phenomena, as mentioned below, and are thus popular as low-dimensional models. These systems do not exhibit the multi-scale dynamics of fully developed three-dimensional turbulence. They are still spatially extended systems, i.e. modelled by partial differential equations, and are thus suitable test cases for the present study~\citep{Holmes2012, Cvitanovic2010}. Below, we briefly describe the model equations and numerical details of the KS and CGL systems. For the chosen parameter values, the two systems exhibit different kind of chaotic behaviour. The KS system exhibits phase turbulence while the CGL system exhibits defect chaos. This qualitative difference in their dynamics may facilitate appropriate generalisation of the results presented in Sections \ref{Res}, \ref{CS} and \ref{SD}.

\subsection{Kuramoto-Sivashinsky system}\label{KS}

The KS system, originally derived for reaction-diffusion processes~\citep{Kuramoto1974} and flame front propagation~\citep{Sivashinsky1977}, models processes driven far from thermodynamic equilibrium by intrinsic long-wavelength instabilities~\citep{Bratanov2013}. The evolution of small perturbation $y(x,t)$ in the one-dimensional KS system is given as,
\begin{equation}\label{KSeq}
\partial_t y = -y\partial_x y - \partial_x^2y - \nu \partial_x^4y, \;\;\; x \in [0,L],
\end{equation}
where $x$ and $t$ are space and time coordinates, respectively, $\nu$ is the viscosity, and the system is $L-$periodic, i.e. $y(x,t) = y(x+L,t)$. 
Equation~\eqref{KSeq} resembles the Navier--Stokes (NS) equations in a few aspects~\citep{Yakhot1981}. In both equations, the energy generation happens at large scales via the linear term $\partial_x^2y$ and energy dissipation happens at small scales via the viscous term (also linear) $\nu \partial_x^4y$. The nonlinear term is energy conserving, it only transfers energy from large to small scales. 

\begin{figure}
 \centerline{\includegraphics[width=1.0\textwidth, trim =0.3cm 0.1cm 2.0cm 0.1cm, clip]{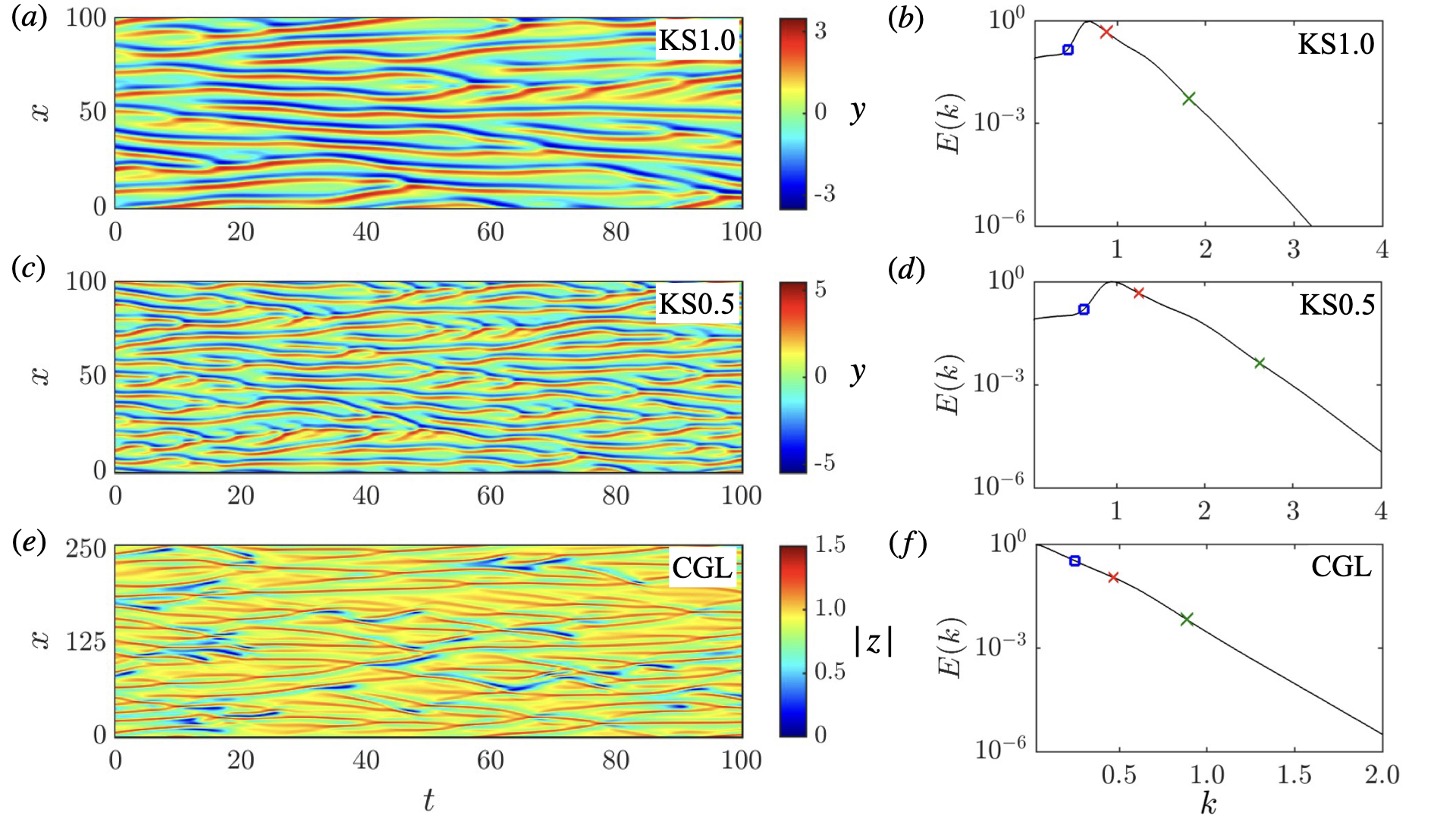}}
 \captionsetup{width=1\linewidth}
  \caption{(a, c, e) Spatiotemporal dynamics and (b, d, f) the normalised energy spectrum ($E(k)$) of the (a, b) KS1.0, (c, d) KS0.5 and (e, f) CGL systems.  The green and red crosses in (b, d, f) correspond to $X_{st}$ at the first and second vertical dashed lines, respectively, in figure~\ref{SPvpt}. The blue squares in (b, d, f) correspond to the cut-off $k$ for chaos synchronisation when information is provided in Fourier domain (see Section~\ref{CS}).}
\label{Sfft}
\end{figure}

The only natural bifurcation parameter in this system is $\tilde{L} = L/(2\pi\nu)$. The system undergoes transition to spatio-temporal chaos at $\tilde{L} \approx 3.66$~\citep{Chate1994}.
In this paper, we consider two KS systems with parameters $(L,\nu)$ = $(32\pi,1.0)$ and $(32\pi,0.5)$. They are referred as KS1.0 and KS0.5, respectively. These systems are numerically solved using ETDRK4 scheme with $N=512$ equispaced points in space and time-step $dt = 0.1$~\citep{Kassam2005}. Figures~\ref{Sfft} (a) and (c) show the space-time evolution of the KS1.0 and KS0.5 systems, respectively, and figures~\ref{Sfft} (b) and (d) show $E(k)$, the energy spectrum normalised by its maximum value, for the two systems. The spectra show that these systems have peak in energy at $k = 1/\sqrt{2\nu}$ and, owing to the hyper-viscous term, a sharp decay in energy after $k \approx 1/\sqrt{\nu}$. It should be noted that this is a log-linear plot, which means that the decay in $E(k)$ with increasing $k$ is sharper than any power-law.

\subsection{Complex Ginzburg--Landau system}

The CGL system is one of the most studied nonlinear system describing a wide range of phenomena in fluid mechanics, condensed matter and string theory~\citep{Aranson2002}.
The one-dimensional CGL system models the spatiotemporal amplitude modulations as,
\begin{equation}
\partial_t z = z + \left(1 + ic_1\right)\partial_x^2z - \left(1-ic_3\right)|z|^2z, \;\;\; x \in [0,L],
\end{equation}
where $z = z_r + iz_i$ is the complex system state with $z_r$ and $z_i$ representing the real and imaginary parts, respectively, $|z|^2 = z_r^2+z_i^2$ and $\left(c_1,c_3\right)$ are the system parameters. 

The system transitions from stable plane wave solutions $(c_1c_3 < 1)$ to phase turbulence (via Benjamin-Fier instability) and then to defect chaos~\citep{Shraiman1992}. We choose the parameters $(c_1,c_3) = (3.50,0.95)$, where the system exhibits defect chaos, and $L=256$. We then numerically solve the system using a fourth-order Runge Kutta scheme with $N=512$ equispaced points in space and time-step $dt = 0.02$. Figures~\ref{Sfft} (e, f) show the space-time evolution and $E(k)$, respectively, of the CGL system. The $|z|=0$ locations in (e) are called the space-time defects. From the spectrum in (f), we note that unlike the KS system, which exhibit a peak at intermediate $k$, energy of the Fourier modes in the CGL system decreases monotonically with increasing $k$.

\section{Methods}\label{Meth}

\begin{figure}
 \centerline{\includegraphics[width=1.0\textwidth, trim =2.8cm 12.1cm 7.0cm 6.1cm, clip]{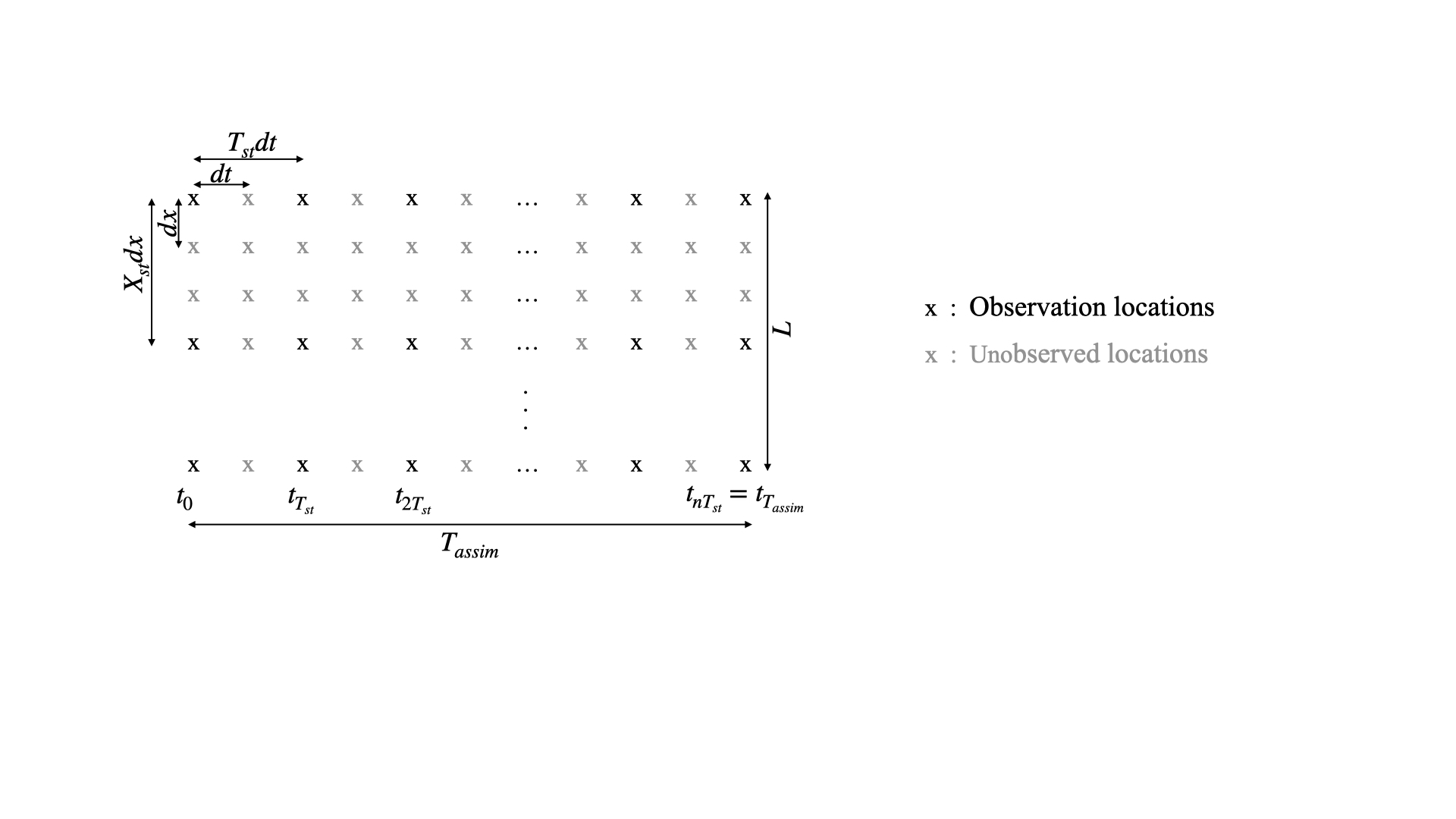}}
 \captionsetup{width=1\linewidth}
  \caption{ Illustration of the computational and observational grids in the DAW for $(X_{st},T_{st})=(3,2)$}
\label{grid}
\end{figure}

The ground truth is obtained from the numerical simulations and thus exist on the computational grid, which has the spacing $dx = L/N$ in space and $dt$ in time. The measurements are sparse in space and time such that the measurements are only known after every $(X_{st},T_{st})$ grid-points in space and time, respectively, as illustrated in figure~\ref{grid}.
The discrete equations for the system's time evolution and the measurement to state relation are denoted as,
\begin{subequations}
\begin{equation}\label{e1}
\mathbf{u}_{j+1} = \mathcal{M}_j\left(\mathbf{u}_j\right),
\end{equation}
\begin{equation}\label{e2}
\mathbf{v}_j  = H\mathbf{u}_j + \epsilon,
\end{equation}
\end{subequations}
where $\mathbf{u}_j$ is the system state at $t = t_{jT_{st}}$,
and the nonlinear evolution operator $\mathcal{M}_j$ is a matrix operator that updates the system state from time $t_{jT_{st}}$ to $t_{(j+1)T_{st}}$. The observation operator $H$ transforms the system state from the ground-truth grid to the observation grid. It is therefore linear in this paper.
The measurement noise ($\epsilon$) is assumed to be uncorrelated in space and time and have Gaussian distribution with zero mean and $\sigma$ standard deviation. The observation error covariance matrix $\mathcal{O}$ is therefore a diagonal matrix of size $m\times m$ with each component equal to $\sigma^2$.
The measurements are available from time $t = 0$ to $T_{assim}$, which is called the data assimilation window (DAW). The number of temporal measurements are given as $n =T_{assim}/T_{st}$.

The three methods used are 4D-Var, EnKF and RC-RNN. The first two are DA methods that use the governing equations as well as the measurements, i.e. information contained in $\mathcal{M}_j$ and $\mathbf{v}_j$. The third method is a model-free ML method that uses the sparse measurements alone, i.e. only information contained in $\mathbf{v}_j$. All three methods implement some form of linear approximations in the time evolution as explained below. Consequently, these methods are not designed to handle large $T_{st}$. We will therefore limit our study to small $T_{st}$.

\subsection{Strong-constraint 4D-Var}\label{M4dV}

4D-Var is a variational method in which the system state is estimated such that it best-fit (in the L2 sense) all the observations in the entire DAW, i.e. it is a smoother~\citep{Carrassi2018}. This is done by minimising the cost function,
\begin{equation}\label{4d1}
J^i = \frac{1}{2}\sum_{j=0}^{j=n} ||\mathbf{v}_j - H \mathbf{u}_j^i ||_{\mathcal{O}^{-1}} + \frac{1}{2}||\mathbf{u}_b - \mathbf{u}_0^i||_{\mathcal{B}^{-1}},
\end{equation}
where $||a||_B = a^TBa$, the superscript $^T$ denotes the conjugate transpose operation, $\mathbf{u}_b$ is a-priori knowledge of the system state at $t_{0}$, which has the error covariance matrix $\mathcal{B}$, and the superscript $i$ indicates the iteration number. In the strong-constraint version, the cost function depends only on the initial condition (i.e. $\mathbf{u}_0^i$), 
\begin{equation}\label{4d2}
J^i = \frac{1}{2}\sum_{j=0}^{j=n} ||\mathbf{v}_j - H M_{0 \rightarrow j}\mathbf{u}_0^i ||_{\mathcal{O}^{-1}} + \frac{1}{2}||\mathbf{u}_b - \mathbf{u}_0^i||_{\mathcal{B}^{-1}},
\end{equation}
where $M_{0 \rightarrow j} = \displaystyle \prod_{l=0}^{j-1}\mathcal{M}_l$ evolves the system state from $t_{0}$ to $t_{jT_{st}}$.
%

%
In 4D-Var, we minimise $J^i$ iteratively by computing the Jacobian $\partial J^i/\partial \mathbf{u}_0^i$. This is a daunting task, which is greatly simplified by using the tangent linear approximation under which $M_{0 \rightarrow j}$ are assumed to be constants. (This linearisation, however, limits $T_{st}$ to be small and $T_{assim}$ to be within the inverse of the maximum Lyapunov exponent~\citep{He2024}.) The linearisation makes the problem quadratic at each iteration. Moreover, adjoint equations are usually formulated for obtaining $\partial J^i/\partial u_0^i$ as,
\begin{equation}\label{adj}
\mathbf{u}_j^{i\dagger} = \mathcal{M}_j^T \mathbf{u}_{j+1}^{i\dagger} + \frac{\partial J^i}{\partial \mathbf{u}_j^i},
\end{equation}
where $\mathbf{u}_j^{i\dagger}$ is the adjoint variable. This adjoint equation is marched backward in time with the initial condition $\mathbf{u}_{n}^{i\dagger} = 0$ and gives $\partial J^i/\partial \mathbf{u}_0^i = \mathbf{u}_0^{i\dagger}$.

We minimise $J$ using a second-order method, which requires calculating the inverse of the Hessian ($\partial^2 J/\partial u_0^2$). In this paper, we do not use the background information, i.e. $\mathcal{B}^{-1}=0$, which usually provides regularisation~\citep{Bauwer2021}. Instead, we use Tikhonov regularisation. Further details on 4D-Var can be found in \citet{Bouttier1999} and \citet{Schlatter2000}, as well as from the codes provided in supplementary material. We note in passing that for turbulent flows at high Reynolds numbers, weak-constraint 4D-Var is a better choice~\citep{Chandramouli2020, He2024}.

\subsection{Ensemble Kalman Filter}\label{Mseq}

EnKF is a sequential method in which the predictions are corrected by passing through a filter whenever an observation is available~\citep{Evensen2003}. 
It uses a Monte-Carlo implementation to calculate the filter by running an ensemble of trajectories of the model dynamics. In practice, only a small size of ensemble is sufficient even for large systems, which makes EnKF an attractive option~\citep{Carrassi2018}.
In DA terminology, the state-vector obtained by forward-marching the governing equations is called the forecast $\mathbf{u}_j^f$ while the corrected estimation is called the analysis $\mathbf{u}_j^a$. The two are related as,
\begin{subequations}
\begin{equation}\label{KF}
\mathbf{u}_j^a = \mathbf{u}_j^f + \mathcal{B}_j^fH^T\left(H \mathcal{B}_j^f H^T + \mathcal{O}_j \right)^{-1} \left(\mathbf{v}_j - H \mathbf{u}_j^f\right),
\end{equation}
\begin{equation}\label{KF2}
\mathcal{B}_j^f = N_e/(N_e-1)\left\langle{\left(\mathbf{u}_j^f - \langle{\mathbf{u}_j^f}\rangle\right) \left(\mathbf{u}_j^f - \langle{\mathbf{u}_j^f}\rangle\right)^T}\right\rangle,
\end{equation}
\end{subequations}
where $\langle . \rangle$ denotes ensemble-averaging and $N_e$ is the number of ensemble members. This correction is based on the Kalman Filter and is thus linear in nature We use the stochastic version of EnKF in which the measurements are also perturbed corresponding to each ensemble member. The measurement error covariance matrix ($\mathcal{O}_j$) is thus also calculated like $\mathcal{B}_j$.
For further details, we refer the reader to~\citet{Evensen2003, Carrassi2018} and~\citet{Pawar2021}, as well as to the codes in supplementary material.

\subsection{Reservoir-computing-based shallow recurrent neural network}\label{RC}

RC-RNN is a purely data-driven ML method designed to predict chaotic systems~\citep{Jaeger2004}. RC-RNN consists of three vector components: input-state ($\mathcal{V}_j$), reservoir-state ($\mathbf{r}_j$), and output-state ($\hat{\mathbf{v}}_j$). They are related to each other as,
\begin{subequations}
\begin{equation}
\mathbf{r}_j = \mathcal{N}\left( A\mathbf{r}_{j-1} + W_{in} \mathcal{V}_j \right),
\end{equation}
\begin{equation}
\hat{\mathbf{v}}_j = W_{out} \mathbf{r}_j,
\end{equation}
\end{subequations}
where the elements of the matrices $A$ and $W_{in}$ are fixed by selecting them from random numbers, $\mathcal{N}$ is an element-wise nonlinear function ($\tanh$ here), and the elements of $W_{out}$ are obtained by minimising the L2-error between $\hat{\mathbf{v}}_j$ and $\mathbf{v}_{j+1}$ during the training phase. The error minimisation is performed via a one-step linear regression, which makes the training of RC-RNN computationally efficient.

In this work, RC-RNN only receives the sparse measurements, $\mathbf{v}_j$, and never sees the full state vector $\mathbf{u}_j$ even in the training phase. The input $\mathcal{V}_j$ consists $(\mathbf{v}_j,\mathbf{v}_j^2)$ for the KS1.0 and KS0.5 systems and $(\mathbf{v}_j, |\mathbf{v}_j|^2\mathbf{v}_j)$ for the CGL system. These choice of $\mathcal{V}_j$ are motivated by the kind of nonlinearity in these systems, which help the network to mimic the system dynamics efficiently. Other alternatives are to use larger reservoir size, which is computationally unfeasible, or to use Gaussian radial basis functions~\citep{Gupta2023}.

Implementation of RC-RNN has three phases: (i) training phase in which a long measurement history is used to fix the elements of $W_{out}$, (ii) initialization phase in which measurements during the DAW (i.e. from $t=0$ to $T_{assim}$) are used to bring the reservoir-state to the system's current state, and (iii) prediction phase in which the network becomes autonomous, i.e. the RC-RNN output at $t_{(j-1)T_{st}}$ is used as the input at $t_{jT_{st}}$. For further details, we refer the reader to~\citet{Gupta2023} and the codes in supplementary material.

\section{Prediction results}\label{Res}


In this section, we first propose a measure of prediction accuracy. Based on that, we obtain the variation in prediction accuracy with increasing spatial sparsity of the measurements. We then discuss the qualitative changes in the predictions for different levels of spatial sparsity.

\subsection{Measure of prediction accuracy: valid prediction time (VPT)}\label{vpt}

We define the normalised root-mean-square-errors calculated on the observation grid as,
\begin{equation}
\mathcal{E}(t;X_{st},T_{st},\sigma) = \frac{\left[\left(\overline{\left(\mathbf{u}^g(t) - \mathbf{u}^a(t)\right)^T\left(\mathbf{u}^g(t) - \mathbf{u}^a(t)\right)}\right)^{0.5}\right]}{S},
\end{equation}
where the superscripts $g$ and $a$ refer to the ground truth and the predicted states, respectively, $S$ is the standard deviation of $\mathbf{u}^g$, the overline indicates the averaging over the observation space, and $\left[ . \right]$ indicates averaging over several repetitions. (At least 200 repetitions are used in all the results.) 
The corresponding error calculated over the ground-truth grid is referred as $\mathcal{E}_{g}$.
Both $\mathcal{E}$ and $\mathcal{E}_{g}$ are functions of time and functional of the measurement parameters $(X_{st},T_{st},\sigma)$. They also depend on the method-specific parameters, such as number of iterations in 4D-Var, number of ensemble members in EnKF and length of the training data in ML. The method-specific parameters for results in each figure are provided in Appendix~\ref{Apar}.

\begin{figure}
 \centerline{\includegraphics[width=0.8\textwidth, trim =0.0cm 5.4cm 3.9cm 2.2cm, clip]{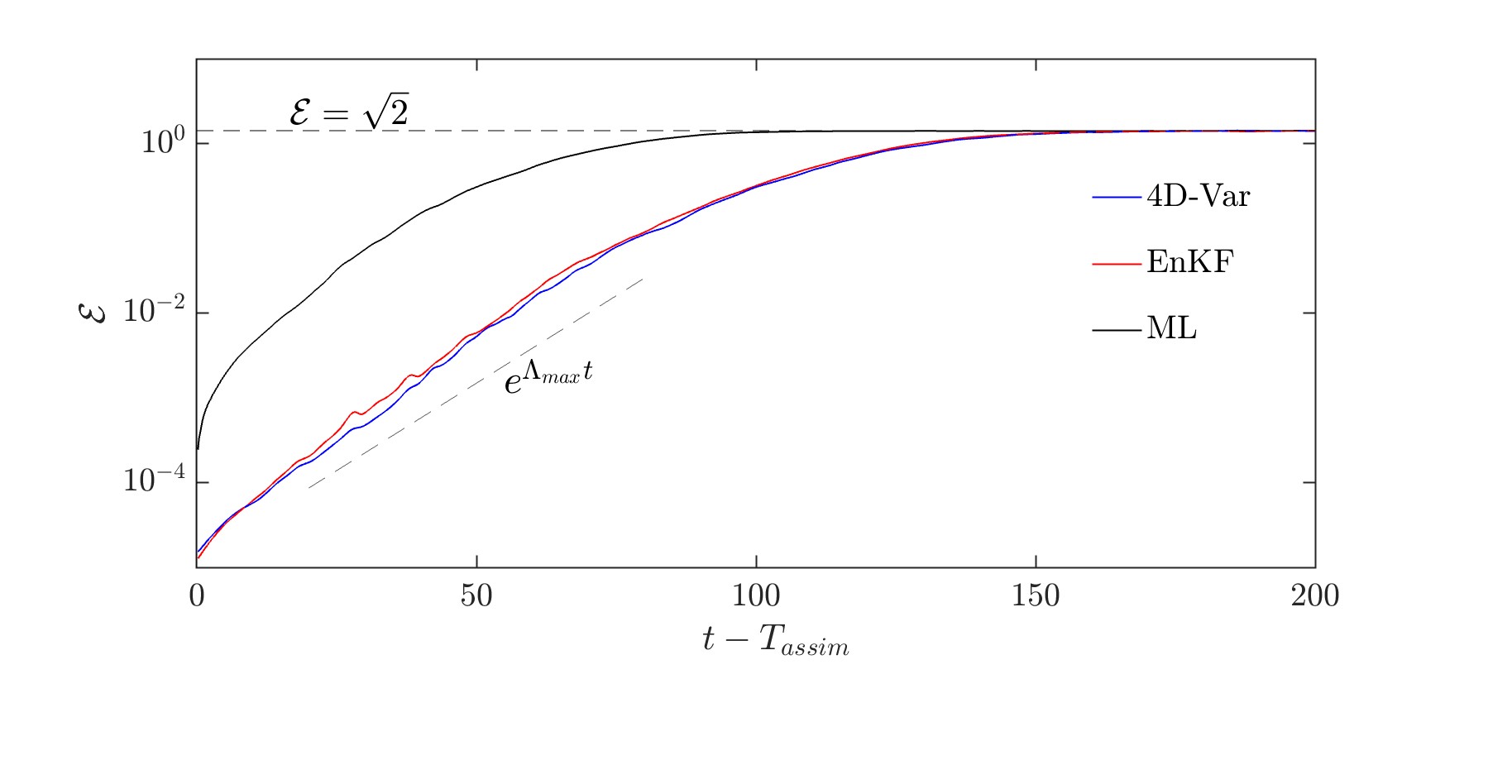}}
 \captionsetup{width=1\linewidth}
  \caption{The evolution of the normalised root-mean-square error for the predictions of KS1.0 system when measurements are available at $(X_{st}, T_{st},\sigma) = (4, 2,1e^{-4}S)$.}
\label{ErK1}
\end{figure}

Figure~\ref{ErK1} shows the evolution of $\mathcal{E}$ for the predictions of KS1.0 system by the three methods (RC-RNN is referred as ML in all the results presented). 
These results show that after an initial transient $\mathcal{E}$ grows approximately as $\exp\left(\Lambda_{max} t \right)$, where $\Lambda_{max}$ is the maximum Lyapunov exponent, and eventually saturates to $\sqrt{2}$. (The saturation to $\sqrt{2}$ suggests that the predictions are completely uncorrelated with the ground-truth).
The faster than $\exp\left(\Lambda_{max} t \right)$ growth of error in the initial transient is because the predicted state at $t = T_{assim}$ is usually not the solution of the governing equation (even for strong-constraint DA methods). This problem is more severe for the ML method because the learned model may also have errors relative to the governing equations.
It is worth noting that $\mathcal{E}$ from the ML method still saturates to $\sqrt{2}$, which indicates that the ML method reproduces the second-order statistics faithfully (see figure~\ref{MLstat} for more details).

We define the prediction accuracy in terms of the time at which $\mathcal{E}$ crosses a threshold. Following~\citet{Pathak2018}, we set the threshold as 0.5 and call the prediction time ($t-T_{assim}$) at which $\mathcal{E}$ crosses 0.5 as valid prediction time ($VPT$). Higher values of $VPT$ therefore indicate higher prediction accuracy.
Although simple, this measure is sufficient to capture the trends in prediction accuracy well. The main advantage of this measure is that it is based on the accuracy of future predictions instead of the accuracy of the reconstruction at the end of the DAW. Therefore, it can also account for possible over-corrections, i.e. minimisation of L2-error at the cost of pushing the system too far away from the attractor, as discussed in Section~\ref{R3}.

\subsection{Prediction accuracy vs spatial resolution}\label{Xres}

\begin{figure}
 \centerline{\includegraphics[width=1\textwidth, trim =0.5cm 8.6cm 0cm 1.9cm, clip]{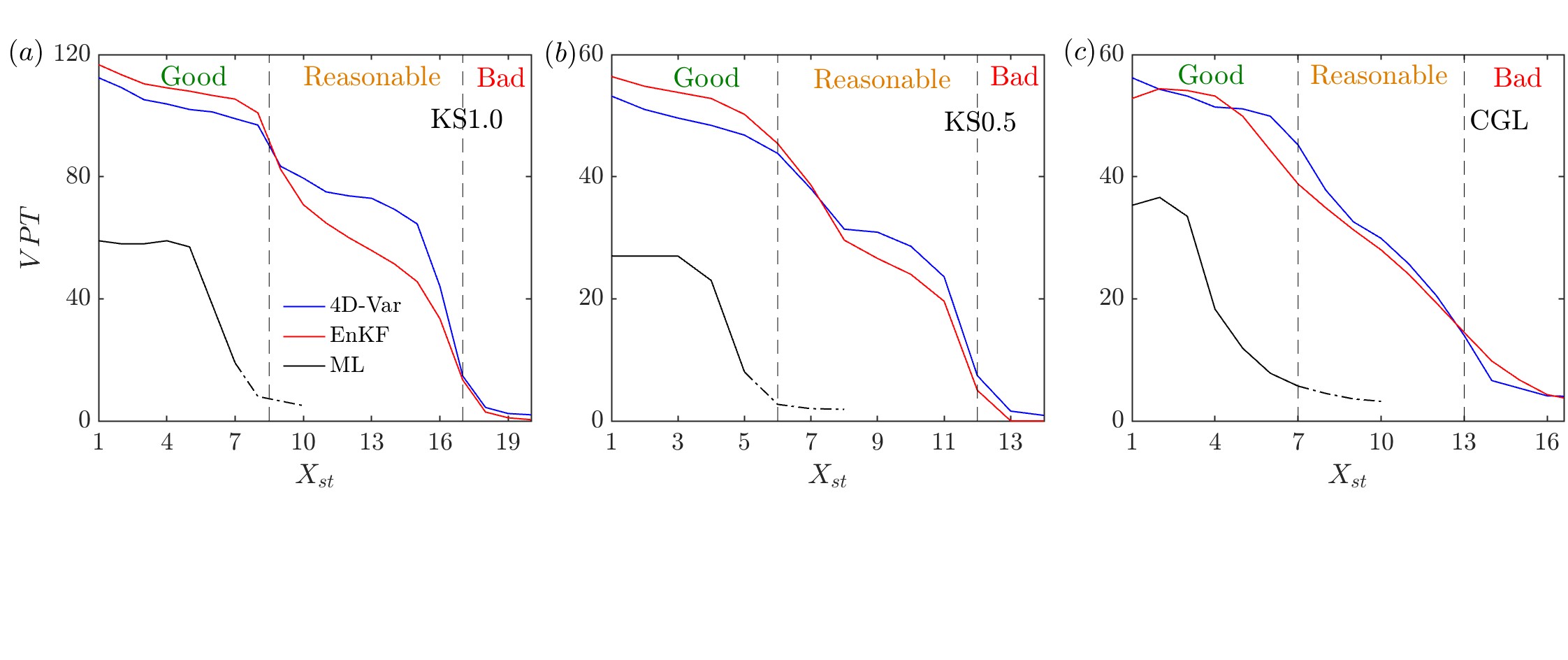}}
 \captionsetup{width=1\linewidth}
  \caption{Variation in $VPT$ with $X_{st}$ for predictions of the (a) KS1.0, (b) KS0.5, and (c) CGL systems using 4D-Var (blue), EnKF (red) and ML (black) methods. The two vertical dashed lines divide the measurement region in three zones, these lines roughly correspond to $X_{st}$ at which $VPT$ steeply changes The dot-dashed line parts of the ML results correspond to statistically incorrect predictions (see figure~\ref{MLstat}).}
\label{SPvpt}
\end{figure}
 
\begin{figure}
 \centerline{\includegraphics[width=1\textwidth, trim =0.4cm 18.4cm 0.4cm 6.3cm, clip]{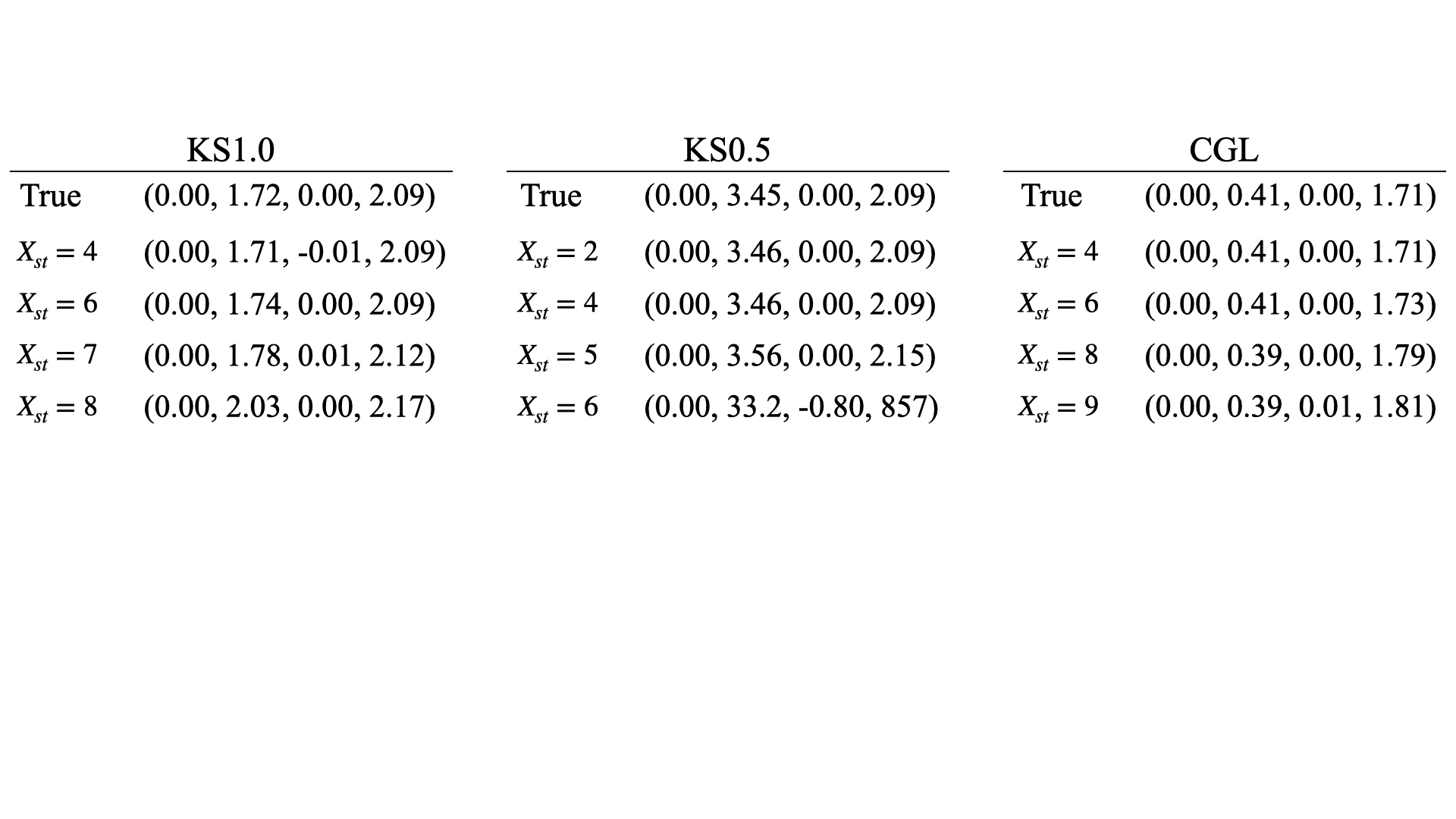}}
 \captionsetup{width=1\linewidth}
  \caption{The first four moments of the true state (top rows) and the predicted states from ML for observations at different levels of spatial sparsity (indicated by $X_{st}$). When the moments from true and predicted states do not match, it indicates that the ML network fails to learn the system's dynamics faithfully.}
\label{MLstat}
\end{figure}

%
Figure~\ref{SPvpt} shows variations in $VPT$ with increasing $X_{st}$ for the (a) KS1.0, (b) KS0.5 and (c) CGL systems. Other measurement parameters are $(T_{st},\sigma) = (2,1.0e^{-4}S)$ for the KS1.0 and KS0.5 systems, $(5,1.0e^{-3}S)$ for the CGL system, and $T_{assim}$ is of the order $\Lambda^{-1}$ of the respective systems (see Appendix \ref{Apar} for method-specific parameters). We observe that $VPT$ decreases with increasing sparsity, i.e. increasing $X_{st}$, for all three prediction methods. The relatively lower values of $VPT$ from the ML method are attributed to the errors in the learned model, while the two DA methods use the exact governing equations. This difference between ML and DA will reduce if we use a larger network or if there are model errors in DA. More concerning matter, however, is that for higher levels of sparsity, the ML method fails to learn the systems' dynamics faithfully. The dot-dashed-line parts of the ML results indicate that corresponding ML predictions are not even statistically correct. Figure~\ref{MLstat} shows the first four moments (mean, variance, skewness and kurtosis) for the ground truth and predictions from ML corresponding to sparse observations. The moments for the predictions are obtained by collecting data from 400 short-term predictions (each for the time $\approx \Lambda^{-1}$). The results show that the predictions from ML become statistically erroneous when the level of sparsity is towards the end of the good-predictions zone (i.e. before the first vertical dashed line).

For all three systems, the two DA methods are able to predict beyond the good-predictions zone but at significantly lower levels of accuracy. At further higher levels of spatial sparsity, except the EnKF results for the CGL system, there is another steep reduction in $VPT$. Further increasing the sparsity of observations does not significantly affect $VPT$.
Based on these steep changes in $VPT$, we heuristically divide the spatial sparsity levels in three zones, i.e. we do not use any quantitative rules or mathematical justification for this division. The three zones are separated by the two dashed vertical lines in figure~\ref{SPvpt}. In the first zone, called good-predictions zone, the measurements are well-resolved and $VPT$ remains close to that for the full resolution measurements. The ML method works only in this zone. In the second zone, called reasonable-predictions zone, $VPT$ from the DA methods reduces significantly as compared to that in the good-predictions zone, but still remains reasonably higher than that in the third zone. In the third zone, called bad-predictions zone, the measurements are sparse and the $VPT$ is reduced to a small fraction of the inverse Lyapunov exponent. Thus, indicating the failure of the DA methods when the measurements are too sparse.

We also note from figure~\ref{SPvpt} that EnKF seems better than 4D-Var in the good-predictions zone and the other way around in the reasonable-predictions zone, particularly in panels (a) and (b). These differences, however, are relatively minor and can be mitigated by changing the method-specific parameters with varying $X_{st}$. 4D-Var method can be improved by adjusting $T_{assim}$ and increasing the number of iterations, while EnKF method becomes more accurate with increasing $T_{assim}$. Therefore, these minor differences are not the focus here.

\subsection{Analysis of the three zones}\label{R3}

All three systems show significant differences in $VPT$ between the three zones. In this section, we analyse qualitative differences in the predictions between these zones to confirm the quantitative results shown in figure~\ref{SPvpt}.

\begin{figure}
 \centerline{\includegraphics[width=1\textwidth, trim =2.0cm 2.8cm 5.5cm 1.6cm, clip]{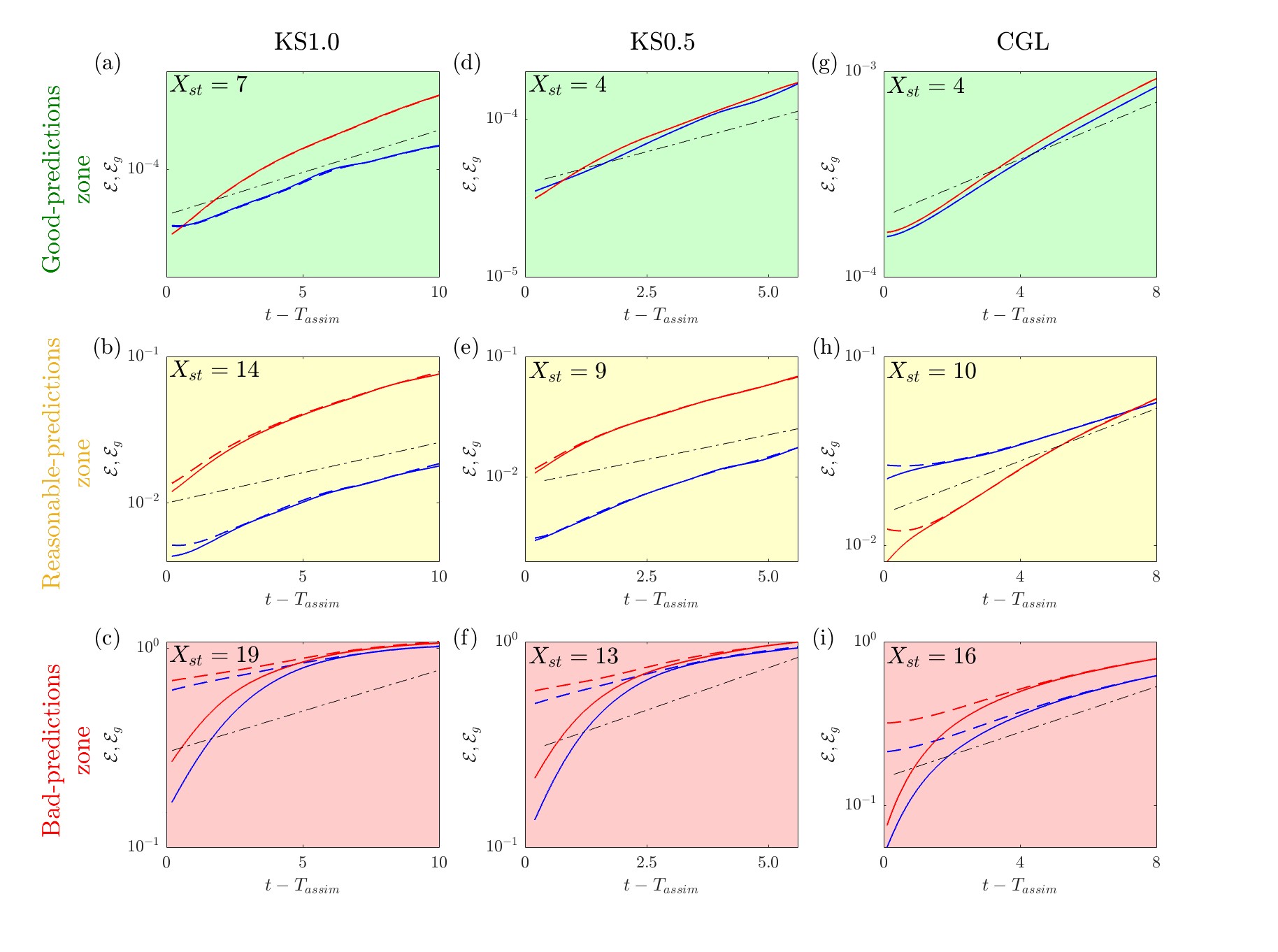}}
 \captionsetup{width=1\linewidth}
  \caption{Time evolution of $\mathcal{E}$ (solid lines) and $\mathcal{E}_g$ (dashed lines) for predictions of the (a-c) KS1.0, (d-f) KS0.5, and (g-i) CGL systems by 4D-Var (blue) and EnKF (red) methods. The dot-dashed lines indicate $\exp(\Lambda_{max}t)$ growth. The difference between the errors on observation grid (solid lines) and ground-truth grid (dashed lines) indicates relatively higher reconstruction error at unobserved locations.}
\label{Err3}
\end{figure}

Figure~\ref{Err3} shows the time evolution of the errors on the observational and ground-truth (computational) grids, $\mathcal{E}$ (solid lines) and $\mathcal{E}_g$ (dashed lines), for predictions of the (a-c) KS1.0, (d-f) KS0.5, and (g-i) CGL systems by 4D-Var (blue) and EnKF (red) methods. The dot-dashed lines indicates growth rate as per $\exp(\Lambda_{max}t)$. The results in the top, middle, and bottom rows correspond to when the spatial resolution is in the good, reasonable, and bad-predictions zones, respectively. The results in the top row show that $\mathcal{E}$ and $\mathcal{E}_g$ are almost indistinguishable and grow approximately as $\exp(\Lambda_{max}t)$. This indicates that the system is predicted with similar accuracy at the observed and unobserved locations. The results in the middle row show that $\mathcal{E}$ and $\mathcal{E}_g$ differ at the beginning of the prediction times. This indicates that the predictions at the unobserved locations are no longer as accurate as those at the observed locations. The growth in $\mathcal{E}$ is similar to $\exp(\Lambda_{max}t)$ while the growth in $\mathcal{E}_g$ is much slower, thus indicating that the additional errors at unobserved locations are not significant. The results in the bottom row show large differences between $\mathcal{E}$ and $\mathcal{E}_g$. The growth in $\mathcal{E}$ is significantly faster than that of $\exp(\Lambda_{max}t)$ while the growth in $\mathcal{E}_g$ is similar to that of $\exp(\Lambda_{max}t)$. This rapid growth in $\mathcal{E}$ indicates that the DA methods over-correct at observation locations. Such over-corrections only minimise the L2-error but the predicted states are pushed too far from the attractor leading to physically inconsistent predictions. We also note here that for the CGL system in the middle row, there is a significant difference in the growth rate of errors for predictions from 4D-Var and EnKF methods (see Appendix~\ref{AppB} for more details).

\begin{figure}
 \centerline{\includegraphics[width=1\textwidth, trim =1.75cm 3.2cm 5.7cm 1.6cm, clip]{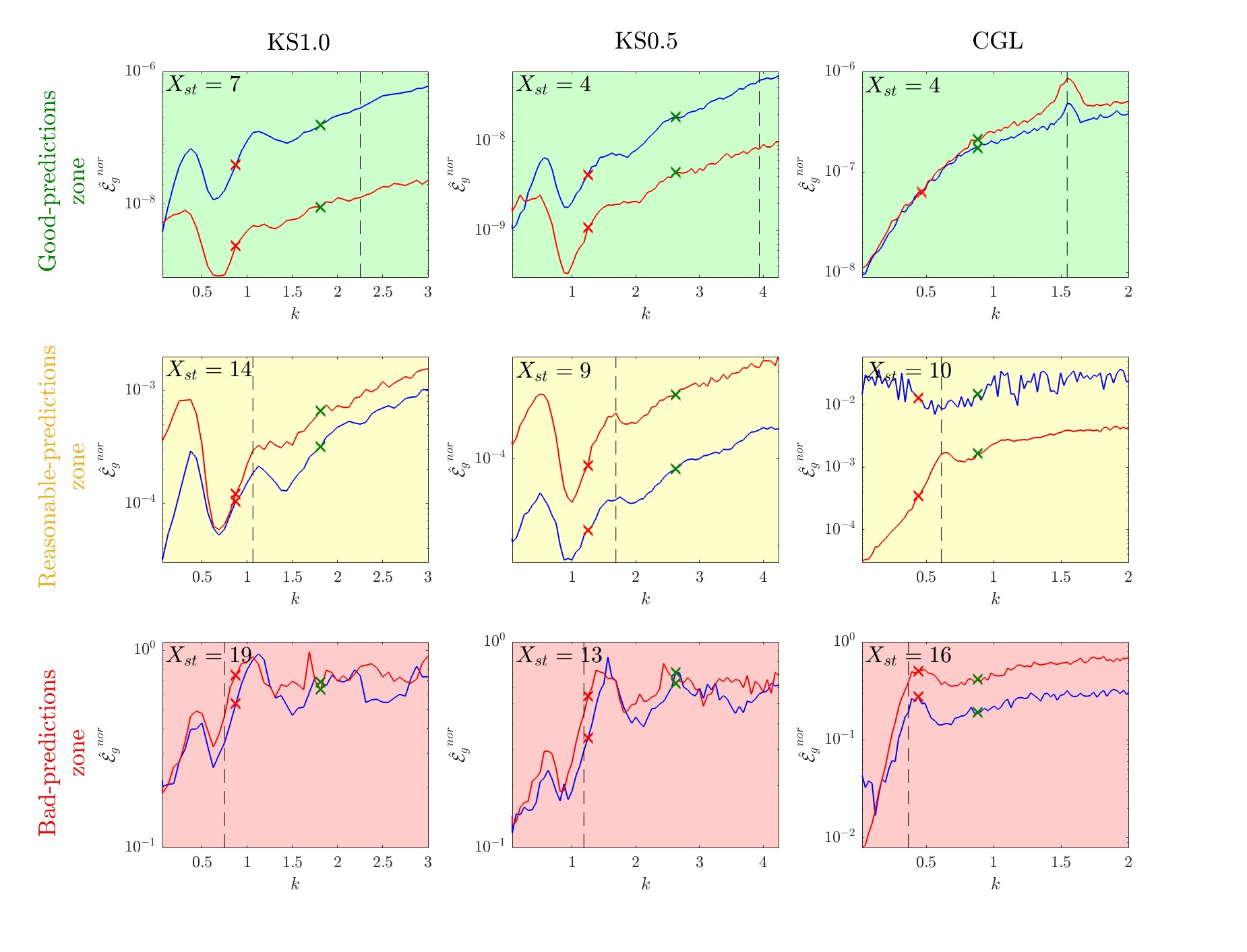}}
 \captionsetup{width=1\linewidth}
  \caption{The normalised error spectrum ($\hat{\mathcal{E}}_g^{nor}$) corresponding to $\mathcal{E}_g$ (at the beginning of the prediction time) shown in the respective panels in figure~\ref{Err3}. The vertical dashed lines indicate the cut-off scales at corresponding $X_{st}$. The green and red crosses correspond to $X_{st}$ at the first and second vertical dashed lines, respectively, in figure~\ref{SPvpt}.}
\label{Err3FFT}
\end{figure}

In figure~\ref{Err3FFT}, we further present the normalised error spectra ($\hat{\mathcal{E}}_g^{nor}$) corresponding to $\mathcal{E}_g$ shown in the respective panels in figure~\ref{Err3}. Specifically, the Fourier transform of $\mathcal{E}_g$ is taken at $t-T_{assim} = T_{st}dt$ (i.e. at the first prediction step) and is normalised by the energy spectrum of the corresponding system (shown in figure~\ref{Sfft}). The blue and red lines correspond to results from 4D-Var and EnKF, respectively, and the vertical dashed lines indicate $k$ up to which the measurements are available for corresponding $X_{st}$. In the top row (good-predictions zone), $\hat{\mathcal{E}}_g^{nor}$ has a similar trend on either side of the dashed line (except for the small bump in the CGL system). This indicates that the predictions of the measured scales are unaffected by the unmeasured scales. Consequently, $VPT$ in this zone does not decrease much with the increasing sparsity levels. In the middle row (reasonable-predictions zone), the overall trends of error are quite similar to that in the top row except for a relative increase in the error in the measured scales. This is most evident from the 4D-Var results for the CGL system, which is further discussed in Appendix~\ref{AppB}. This indicates that the prediction of the measured scales is affected by the unmeasured scales. Consequently, $VPT$ in this zone is significantly lower than that in the good-predictions zone. In the bottom row (bad-predictions zone), the errors in the measured scales are much lower relative to the errors in the unmeasured scales (comparing with the results in the top and middle rows). This indicates the over-correction of the measured scales. The errors in the unmeasured scales, however, drastically increase the overall errors and reduce $VPT$ in this zone as seen in figures~\ref{SPvpt} and~\ref{Err3}.

\section{Threshold sparsity level for chaos synchronisation and DA predictions}\label{CS}

In this section, we relate the sparsity levels up to which the DA methods can predict, i.e. the second vertical dashed lines in figure~\ref{SPvpt}, with the threshold sparsity levels up to which chaos synchronisation can be achieved.
Chaos synchronisation is defined as a process in which two coupled chaotic systems, which have different states, adjust such that their motion eventually exhibit a common behaviour~\citep{Boccaletti2002}. \citet{Kocarev1997} showed that two unidirectionally coupled spatiotemporally chaotic systems can be synchronised using coupling at finite number of spatial points. This suggests that only scales larger than some threshold level govern the dynamics of spatiotemporally chaotic systems. \citet{Yoshida2005} and \citet{Lalescu2013}, from different perspectives, studied chaos synchronisation of turbulent flows. They showed that small scales in homogeneous isotropic turbulence can be reconstructed to machine accuracy by using information available in the large scales alone.

We follow \citet{Yoshida2005} to find the threshold spatial sparsity level up to which the information is required to achieve chaos synchronisation. First, we define two systems obeying the identical governing equations and boundary conditions. We refer to them as master and slave systems whose state at time $t$ is $\mathbf{u}^{(m)}_t$ and $\mathbf{u}^{(s)}_t$, respectively.  The master system is simulated independently for which we have coarse-grained observations, with sparsity parameter $X_{st}$, available continuously in time (i.e. $T_{st}=1$). The slave system, which is uncorrelated from the master system at $t=0$, is evolved such that $\mathbf{u}^{(s)}_t$ is replaced by $\mathbf{u}^{(m)}_t$ at the observation locations at every time instant. Consequently, the dynamics of the slave system at large-scales is coupled to that of the master system. To find if such a coupling will lead to synchronisation of the two systems, we calculate the difference between their states as $S_{err}(t) = \left[|\mathbf{u}^{(m)}_t - \mathbf{u}^{(s)}_t|^2\right]$. Chaos synchronisation is confirmed when $S_{err}$ reduces with time such that it eventually converges to 0 (to the machine precision).

\begin{figure}
 \centerline{\includegraphics[width=1\textwidth, trim =6.25cm 1.85cm 7.2cm 1.35cm, clip]{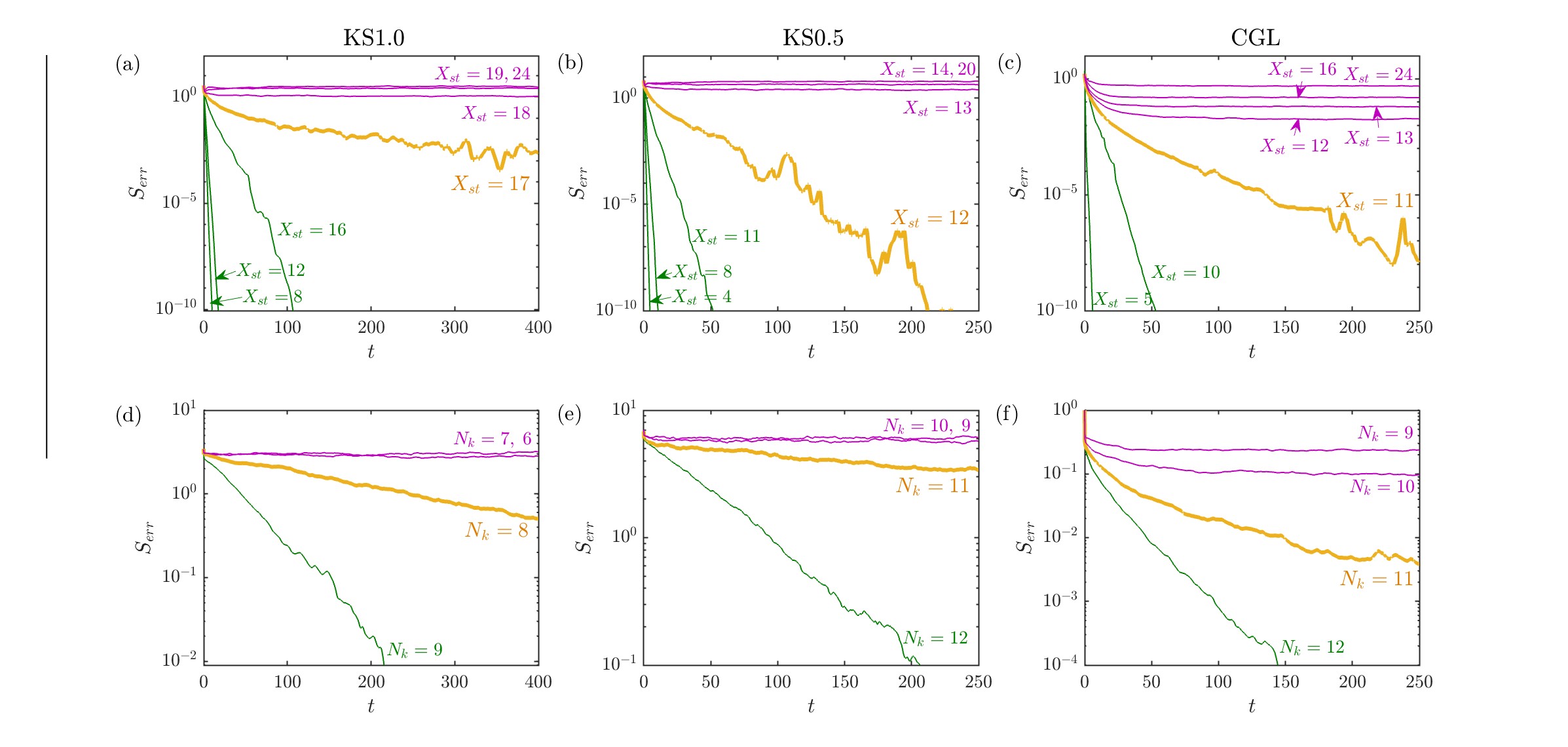}}
  \captionsetup{width=1\linewidth}
  \caption{The average mean-square distance ($S_{err}$) between the states of the master and slave (a, d) KS1.0, (b, e) KS0.5, and (c, f) CGL systems. The slave system is driven by spatially sparse observations (top row) or first $N_k$ Fourier modes (bottom rows) of the master system. Reduction in $S_{err}$ with time indicates that the two systems will be synchronised after a sufficiently long time. The thick orange lines correspond to the threshold sparsity levels up to which synchronisation is achieved, while the green and purple lines correspond to denser and sparser observations than the threshold level.}
\label{csf}
\end{figure}

Figure~\ref{csf} (top row) shows the evolution of $S_{err}$ at various sparsity levels for the (a) KS1.0, (b) KS0.5, and (c) CGL systems. The reduction in $S_{err}$ with $t$ indicates that the slave system will be synchronised with the master system after a sufficiently long time. We find that the threshold sparsity levels (i.e. the coarsest resolution) for synchronisation of the KS1.0, KS0.5, and CGL systems are $X_{st} = 17,$ 12, and 11, respectively. The threshold conditions for the KS1.0 and KS0.5 systems are exactly the same $X_{st}$ up to which the DA methods work (see figure~\ref{SPvpt}). The threshold condition for the CGL system is slightly lower $X_{st}$ (i.e. slightly higher resolution) than the one ($X_{st}=13$) up to which the DA methods work.  
The ability of the DA methods to skilfully predict the CGL system for $X_{st} > 11$ is related to the observation that the slave system is still partially correlated with the master system at $X_{st} > 11$ (figure~\ref{csf} (c)). A plausible reason for this difference in the KS systems and CGL system could be because of the difference seen in their spectra. The KS systems have a dominant mode at an intermediate $k$ while in the CGL system the energy decreases monotonically with $k$.
However, as seen in figure~\ref{SPvpt}, the prediction accuracy of the DA methods for the CGL system falls rapidly after $X_{st} = 11$ and the DA methods fail for $X_{st}>13$. We therefore conclude that the spatial resolution up to which the DA methods can skilfully predict is nearly the same as the threshold resolution up to which chaos synchronisation can be achieved.

We also perform chaos synchronisation in the Fourier space by substituting the first $N_k$ Fourier modes (i.e. largest scales) from the master system in the slave system.
The bottom row of figure~\ref{csf} shows that chaos synchronisation of KS1.0, KS0.5, and CGL systems are achieved when the slave system is driven by the first $N_k = 8,$ 11, and 11 Fourier modes of the master system. There are two points to note from these results. First, they show the sharpness of the threshold condition for chaos synchronisation. This is evident from the rapid change in the decay rate of $S_{err}$ for change in $N_k$ near the threshold level. Second, the wavenumber up to which the measurements are required for chaos synchronisation are marked by blue squares in figure~\ref{Sfft}. It shows that much fewer measurements are required to achieve chaos synchronisation when the measurements are available in the Fourier domain. This is similar to the DA results reported in~\citet{DiLeoni2020}.

\section{Relation between the prediction accuracy and system dynamics}\label{SD}

Predictions from both DA methods for all three systems suffer from sudden decrease in the prediction accuracy as the sparsity levels change from the good to reasonable-predictions zone. Both DA methods then fail in the bad-predictions zone, while the ML method gives physically consistent predictions only in the good-predictions zone. Therefore, we hypothesise that the division in three sparsity level zones, which was done heuristically in Section~\ref{Xres}, is related to the system dynamics. This means that the division is likely to be independent of the DA and ML methods used. It may be tempting to explain the different zones directly by looking at the energy spectra in figure~\ref{Sfft}. The green and red markers represent the cut-off scales corresponding to $X_{st}$ which divide the sparsity level zones in figure~\ref{SPvpt}. The green marker is in the rapidly decaying region and the red marker is in the energetic modes region. However, there are two major problems with such an explanation. First, if the information is directly provided in the Fourier domain then fewer wavenumbers are required to achieve the same level of prediction accuracy. The blue squares in figure~\ref{Sfft} show the cut-off wavenumber required for chaos synchronisation if the information is provided directly in the Fourier space (see Section~\ref{CS}). Second, and somewhat related to the first point, such an explanation based on the dominant length scale is merely a guess work with no guarantee of generalisation (see figure 8.6 and related discussion in \citet{Holmes2012}). In this section, we therefore look at several measures of the system dynamics for plausible explanation of the results shown in Sections~\ref{Res} and~\ref{CS}.

\subsection{Two-point linear and nonlinear correlations}\label{acf}

\begin{figure}
 \centerline{\includegraphics[width=1\textwidth, trim =0.45cm 0.3cm 1.025cm 0.7cm, clip]{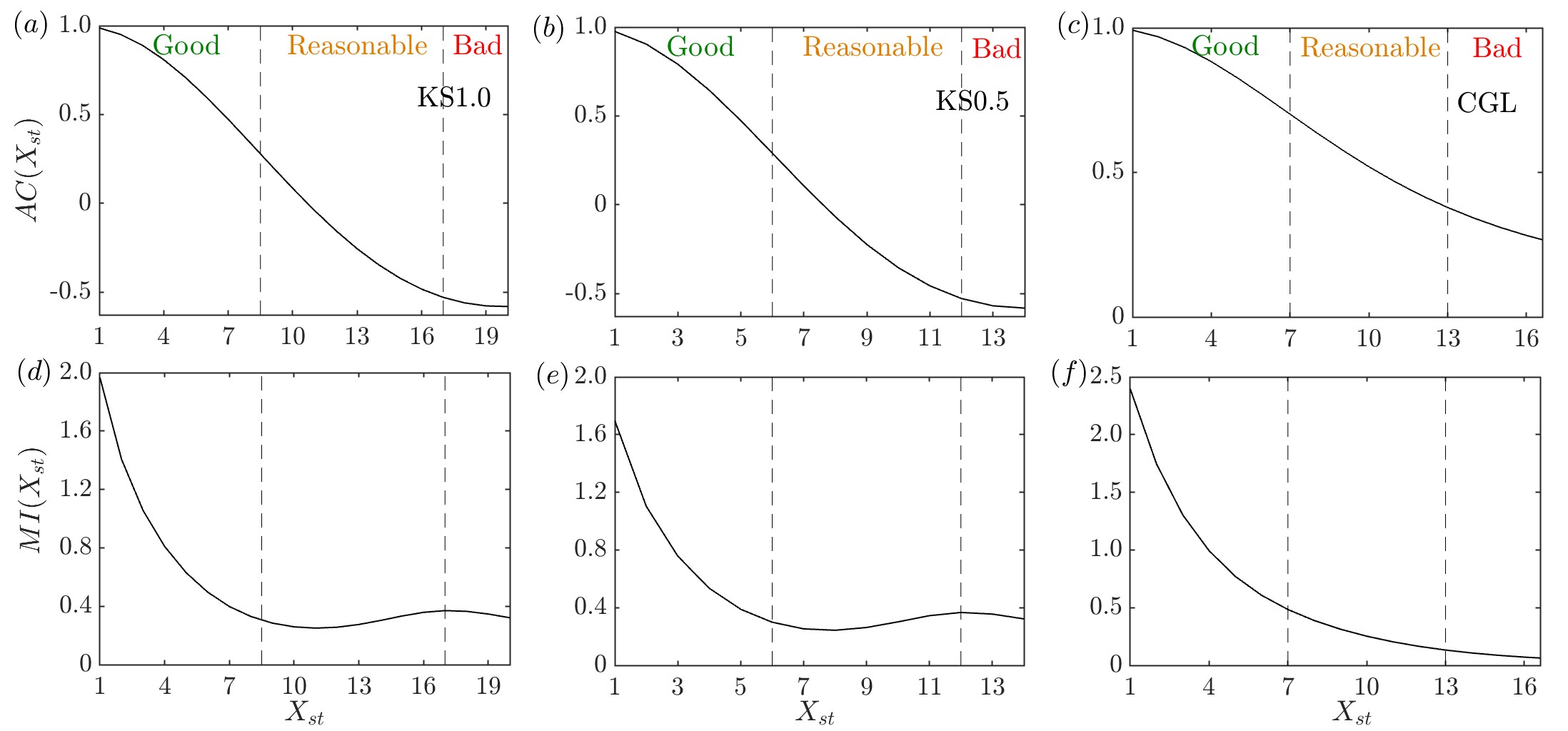}}
 \captionsetup{width=1\linewidth}
  \caption{The (a-c) autocorrelation $AC(X_{st})$ and (d-f) mutual information $MI(X_{st})$ between two points at distance $X_{st}dx$ for the (a, d) KS1.0, (b, e) KS0.5, and (c, f) CGL systems. There are no consistent qualitative or quantitative changes in $AC$ and $MI$ that coincide with the vertical dashed lines for these systems.}
\label{DC}
\end{figure}

If two random variables are correlated to each other, observing one variable then can provide a reasonable estimation of the other.
The ability to reconstruct turbulence from spatially sparse measurements is therefore often compared with the correlation between the system state at observed and unobserved locations~\citep{Li2019, Wang2021}.
In figure~\ref{DC}, we also present variations in the (a-c) autocorrelation ($AC(X_{st})$) and (c) mutual information ($MI(X_{st})$) between two points at distance $X_{st}dx$ for the (a, d) KS1.0, (b, e) KS0.5 and (c, f) CGL systems.
$AC$ measures the linear dependence between two spatially separated points and is defined as,
\begin{equation}
AC(X_{st}) =  \frac{Cov(u(x),u(x+X_{st}dx))}{Var(u(x))},
\end{equation}
where $Cov$ stands for the covariance and $Var$ stands for the variance. We use MATLAB function corr to calculate $AC(X_{st})$. Higher values of $AC$ should lead to higher prediction accuracy. However, DA methods can give accurate predictions even for low values of $AC$ as shown in \citet{Wang2021}. We find that $AC$ is not indicative of the prediction accuracy by the DA methods for the systems studied in this paper.

Unlike $AC$, $MI$ is non-negative and not limited to linear dependence. $MI$ measures the information gained of one random variable by observing another. It is defined as,
\begin{equation}
MI(X_{st}) = \sum P\left(u(x),u(x+X_{st}dx)\right) \log \left( \frac{P\left(u(x),u(x+X_{st}dx)\right)}{P\left(u(x)\right)^2} \right),
\end{equation}
where $P\left(u(x),u(x+X_{st}dx)\right)$ is the joint probability distribution of the spatially separated signals, $P\left(u(x)\right)$ is the probability distribution of the signal and $\sum$ represents summation over all the combination of values of $u(x)$ and $u(x+X_{st}dx)$.
We calculate $MI(X_{st})$ by using 25 bins for the KS1.0 and KS0.5 systems and 15 bins for the CGL system, and we use log with base 2 (i.e. the unit of MI is in bits).
For the KS1.0 and KS0.5 systems, the second dashed lines coincide with the local maxima in $MI$. However, this criterion does not hold for the CGL system and thus remains inconclusive. We therefore conclude that two-point correlations do not explain the prediction accuracy of the DA methods. This result may relate to the fact that the correlation length is not a good measure of the dynamics of chaotic systems~\citep{Egolf1994}.


\subsection{Conditional correlation dimension}

Although weakly turbulent systems are modelled by partial differential equations, i.e. they are infinite-dimensional, the actual dynamics evolve on a finite dimensional inertial manifold. The correlation dimension ($C_d^*$), proposed by~\citet{Grassberger1983}, is a measure of the dimensionality of such manifolds. It is calculated by first measuring the fraction of times ($C_r$) the system comes within a distance $r$ of any point on the system's trajectory and then calculating the log-log slope of $C_r$-$r$ curve for diminishingly small $r$, i.e.
\begin{equation}
C_d^* = \lim_{r \to 0}\frac{\log(C_r)}{\log(r)}.
\end{equation}
The calculation of $C_r$ can be carried out using sparse measurements or even scalar measurements (by augmenting them with the time-delayed measurements). The precise calculation of $C_d^*$ from $C_r-r$ curve requires noise-free long time-series data, whose length increases exponentially with $C_d^*$. The original purpose of $C_d^*$ is for differentiating between chaotic, non-chaotic and stochastic systems. We are interested in only knowing the variation in the system's complexity captured by the sparse measurements as $X_{st}$ changes. We, therefore, make two simplifications in the calculation of the manifold dimension. First simplification is a condition that we will only use the spatially sparse observation data to calculate $C_r$, i.e. no augmentation with time-delayed measurements is performed. Second simplification is an approximation that we do not attempt to reach $r \to 0$ regime as long as a sufficiently linear tail is obtained. We therefore refer to this measure as the approximate conditional correlation dimension ($C_d$).


\begin{figure}
\centering
  \begin{subcaptiongroup}
  \centering
    \parbox[b]{1\textwidth}{%
    \centering
    \includegraphics[width=1\textwidth, trim =0.1cm 8.7cm 0.0cm 1.85cm, clip]{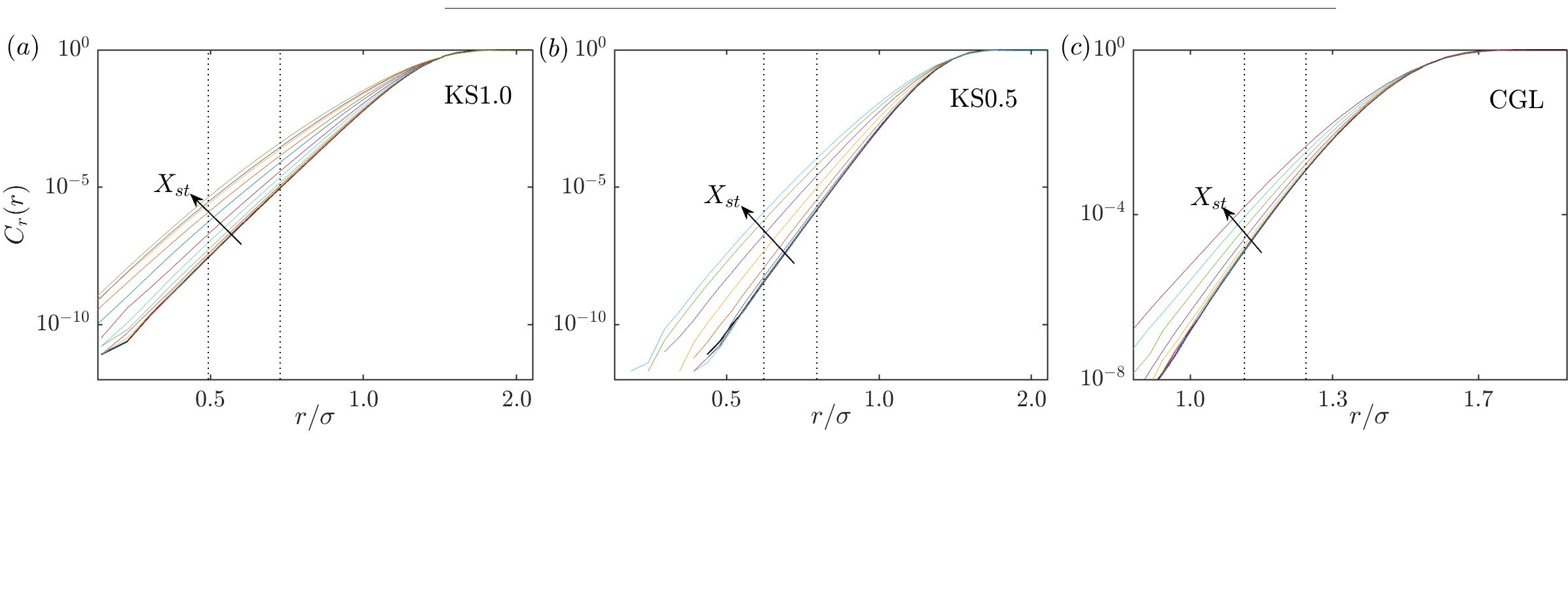}}
   \parbox[b]{1\textwidth}{%
    \centering
    \includegraphics[width=1\textwidth, trim =0.3cm 8.25cm 0.75cm 2.1cm, clip]{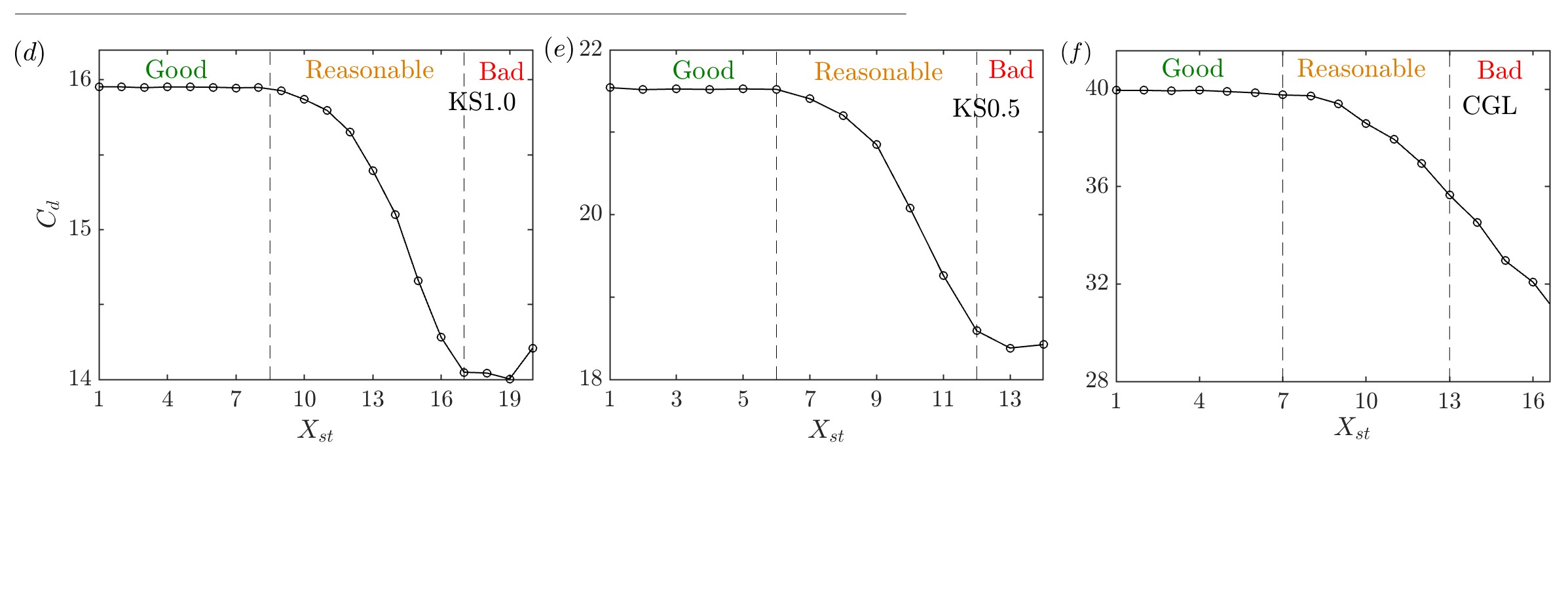}}
\end{subcaptiongroup}
 \captionsetup{width=1\linewidth}
  \caption{The log-log plots of $C_r$ vs $r$ for the (a) KS1.0, (b) KS0.5, and (c) CGL systems. The variations in $C_d$ (calculated from the $C_r-r$ slope between the dotted lines) with $X_{st}$ for the (d) KS1.0, (e) KS0.5, and (f) CGL systems. $C_d$ remains unchanged within the good-predictions zone.}
\label{Cr}
\end{figure}

The top row of figure~\ref{Cr} shows the log-log plots of $C_r$ vs $r$ for the (a) KS1.0, (b) KS0.5 and (c) CGL systems at continuously varying values of $X_{st}$. As the measurements get sparser, i.e. $X_{st}$ increases, the slope of $C_r-r$ decreases. This is because as the system state is projected on increasingly smaller observation space the trajectories come closer to each other.
$C_d$ is calculated from the $C_r-r$ slope between the two vertical dotted lines. The bottom row of figure~\ref{Cr} shows the variation in $C_d$ with $X_{st}$ for the three systems. At full resolution, i.e. $X_{st}=1$, the value of $C_d$ for the KS1.0 and KS0.5 systems is approximately two times the $N_k$ required for chaos synchronisation (see figure~\ref{csf}). The factor of two is because of the Nyquist--Shannon sampling theorem. For the CGL system, $C_d$ is four times the $N_k$ required for chaos synchronisation. The another factor of two is to account for the complex state vector. This shows that $C_d$ is able to capture the dimension of the manifold on which these systems are evolving.

We observe that $C_d$ remains almost unchanged in the good-predictions zone and start to decrease as the measurements get further sparser in the reasonable-prediction zone. We, therefore, conclude the good-predictions zone to be in which the sparse observations can still capture the full complexity, as measured by $C_d$, of the system dynamics. We also conclude this to be the condition for the shallow ML network used in this study to work. For the transition from reasonable-to bad-predictions zone, we find that $C_d$ for the KS1.0 and KS0.5 systems show a change in trend, it either plateaus or even increase in the bad-predictions zone. However, such clear qualitative change in the trend is not seen for the CGL system. $C_d$ thus conclusively explain only the transition from the good to reasonable-predictions zone, but not for the reasonable to bad-predictions zone.

\subsection{Conditional entropy}\label{IG}

We aim to quantify the contribution of the sparse observations in producing the future state of the system to understand the variation in $VPT$ with $X_{st}$. Towards that purpose, we first introduce the Shannon entropy ($H$) as a measure of information in the system state~\citep{Shannon1948}. It is defined for a $p$-dimensional state-vector $\mathbf{w} = (w_1,w_2,\cdots ,w_p)$ as,
\begin{equation}\label{shan}
H\left(\mathbf{w}\right) = \sum_{w_1,w_2,\cdots ,w_p} - \rho\left(w_1,w_2,\cdots ,w_p\right) \log \left(\rho\left(w_1,w_2,\cdots ,w_p\right) \right),
\end{equation}
where $\rho\left(w_1,w_2,\cdots ,w_p\right)$ is the joint probability distribution on all possible pairs of the elements of $\mathbf{w}$, and $\log$ is calculated with base 2 (i.e. the unit of $H$ is in bits). If no other information (or observations) are available then $H$ denotes the uncertainty in determining the system state~\citep{Cover2006}. When other processes or past states ($\tilde{\mathbf{w}}$) are observed, the uncertainty is reduced~\citep{Lozano2022}. The conditional entropy ($H\left(\mathbf{w}|\tilde{\mathbf{w}}\right)$) quantifies the uncertainty in determining $\mathbf{w}$ when $\tilde{\mathbf{w}}$ is known and is given as,
\begin{equation}\label{Cshan}
H\left(\mathbf{w}|\tilde{\mathbf{w}}\right) = \sum_{\mathbf{w},\tilde{\mathbf{w}}} - \rho\left(\mathbf{w},\tilde{\mathbf{w}}\right) \log \left(\frac{\rho\left(\mathbf{w},\tilde{\mathbf{w}}\right)}{\rho\left(\tilde{\mathbf{w}}\right)} \right).
\end{equation}

Ideally, we want to calculate the uncertainty in the future state $\mathbf{u}$ (at $t > T_{assim}$) given all the past measurements ($\mathbf{v}$) (from $t = 0$ to $T_{assim}$).
However, consider that each component of $\mathbf{u}$ can be in $N_b$ discrete intervals, where $N_b$ is the number of bins in which the data values are divided, then the total number of bins for the joint probability will be $N_b^N$.
It is obvious that such calculations cannot be performed even for systems of moderate size ($N ~ O(10^1)$). This is similar to the problem of propagating the Fokker-Planck equations of the probability density functions for nonlinear systems~\citep{Colburn2011}.
We therefore make simplifications and calculate two versions of $H\left(\mathbf{w}|\tilde{\mathbf{w}}\right)$. The first version is based on exploiting the spatial homogeneity of the system and locality of the interactions~\citep{Pathak2018}. In this version, we calculate the uncertainty in determining the system state at location $x$ and time $t_{(k+1)T_{st}}$ (referred as $v_{k+1}(x)$) from a single-instant past measurements at locations $x-X_{st}$, $x$ and $x+X_{st}$ and time $t_{kT_{st}}$ (referred as $v_{k,{X_{st}}}(x)$). The first version of the conditional entropy is therefore referred as $H\left(v_{k+1}(x)|v_{k,X_{st}}(x)\right)$. The second version is based on compressing the information contained in the observed state in terms of the spatial mean. The second version of conditional entropy is therefore referred as $H\left(\overline{\mathbf{v}}_{k+1}|\overline{\mathbf{v}}_{k}\right)$. Other choices for data compression could be the principal components of the observed states or autoencoder-based features, but we do not considered them here.

\begin{figure}
 \centerline{\includegraphics[width=0.95\textwidth, trim =1.25cm 2.55cm 0.0cm 0.7cm, clip]{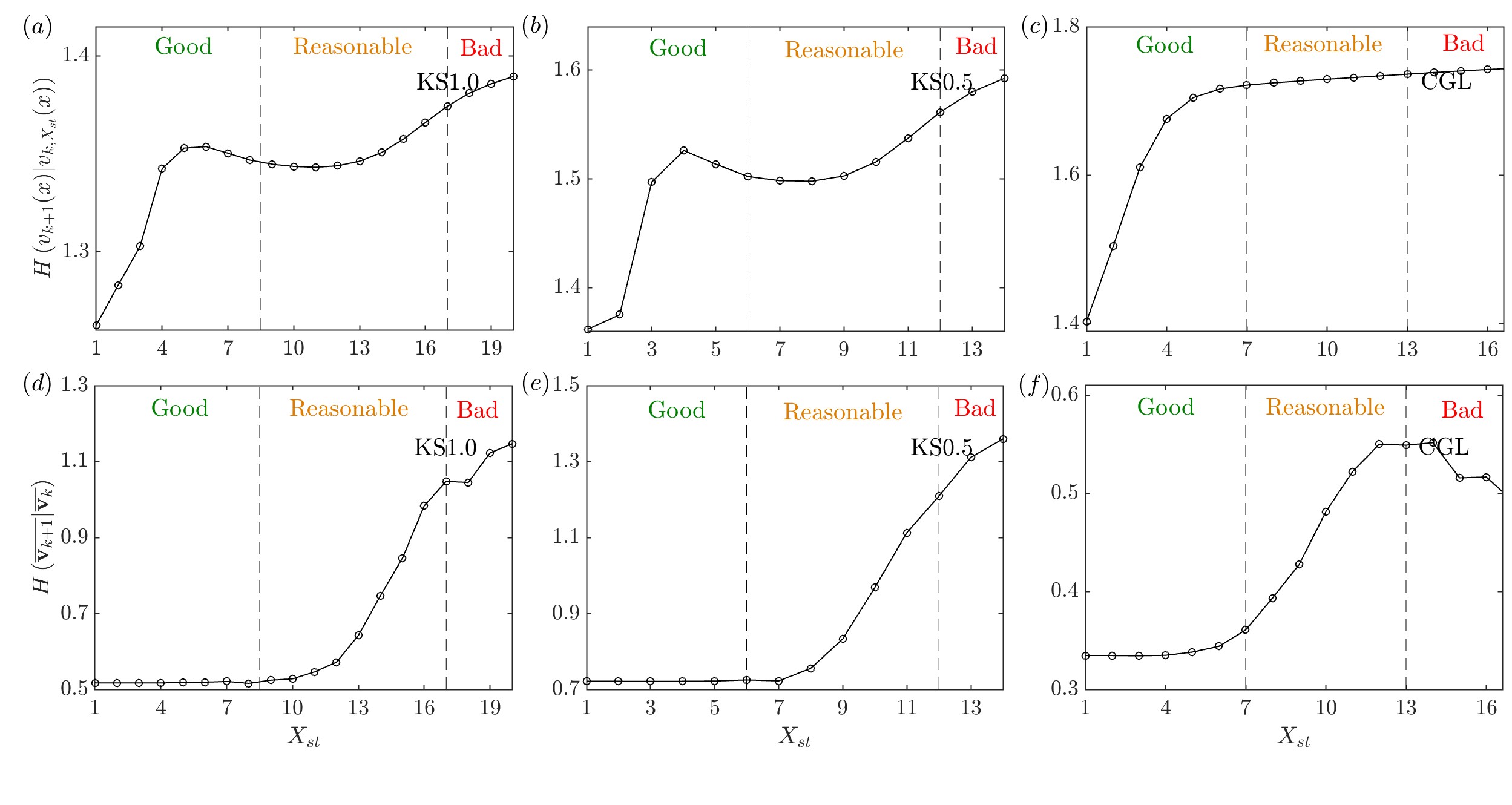}}
 \captionsetup{width=1\linewidth}
  \caption{The variation in the two versions of conditional entropy (a-c) $H\left(v_{k+1}(x)|v_{k,X_{st}}(x)\right)$ and (d-f) $H\left(\overline{\mathbf{v}}_{k+1}|\overline{\mathbf{v}}_{k}\right)$ with $X_{st}$ for the (a, d) KS1.0, (b, e) KS0.5, and (c, f) CGL systems. The first version (top row) fails to explain the division in to zones, but the second version (bottom row) explains the division.}
\label{Hu}
\end{figure}

Figure~\ref{Hu} shows (a-c) $H\left(v_{k+1}(x)|v_{k,X_{st}}(x)\right)$ and (d-f) $H\left(\overline{\mathbf{v}}_{k+1}|\overline{\mathbf{v}}_{k}\right)$ for the (a, d) KS1.0, (b, e) KS0.5, and (c, f) CGL systems. The top row shows that $H\left(v_{k+1}(x)|v_{k,X_{st}}(x)\right)$ exhibits qualitative changes with varying $X_{st}$, but these changes do not happen close to the vertical dashed lines. We, however, cannot conclude if the conditional entropy is not able to explain the predictability or if our simplified version is not good. The second version in the bottom row shows that the qualitative changes in $H\left(\overline{\mathbf{v}}_{k+1}|\overline{\mathbf{v}}_{k}\right)$ with varying $X_{st}$ happen close to the vertical dashed lines. We therefore conclude that the measures from information theory can explain the prediction accuracy of DA methods but such measures must be defined well.

\section{Conclusion}\label{Con}

We use data-driven methods, DA and ML, to predict two weakly turbulent systems from spatially sparse observations. We choose two popular DA methods, 4D-Var and EnKF, which are significantly different from each other. The ML method, RC-RNN, is model-free and uses only spatially sparse measurements. The two chosen systems, KS and CGL, also exhibit qualitatively different chaotic dynamics. With this diverse set of methods and systems, we analyse the effect of spatial sparsity levels on the prediction accuracy. We ask three research questions - the sparsity level up to which the data-driven methods work, variation in their performance with sparsity level, and if the predictive performance of data-driven methods can be explained in terms of the system's dynamics.

We find that the sparsity level up to which DA methods work is almost the same as the threshold sparsity level up to which chaos synchronisation of the corresponding systems can be achieved. The ML method requires even higher resolution. The main implication of this finding is that there is a firm limit on the spatial sparsity level, governed by the system's dynamics, up to which the application of data-driven methods is meaningful. This also confirms the results in \citet{Li2019} who reported that the reconstruction of Kolmogorov flows using 4D-Var is only successful when the resolution is around the threshold level required for chaos synchronisation. Similarly, \citet{Suzuki2017} found EnKF to perform no better than correlation-based estimation when measurements are too sparse.

Within this threshold sparsity level, the prediction accuracy shows interesting variations with sparsity level. In high-resolution good-predictions zone, the prediction accuracy of the DA methods remains almost as good as for full-resolution observations. The ML methods successfully predict only in the good-predictions zone. On further increasing the sparsity level, we enter the reasonable-predictions zone in which the DA methods are still able to predict but with reduced accuracy. We explain the good-predictions zone in terms of the system's dynamics by using the correlation dimension, which measures the dimension of the inner manifold on which the system evolves. In the good-predictions zone, the observations remain dense enough to accurately capture the fractal manifold of the system's dynamics.
The main implication of this finding is that it provides the framework to determine if further increasing the resolution will result in improved performance.
The model-free ML method works only in this zone. Thus, this finding also informs the sparsity level up to which the application of model-free ML methods is meaningful, and beyond which the governing equations must be incorporated.

These results are obtained for weakly turbulent systems and there is no rigorous theory to assume that they will hold for fully turbulent flows. There is empirical evidence, however, for a cautious extension of these results. For example, the relation between the spatial resolution for successful prediction by DA and for chaos synchronisation is reported for fully turbulent flows~\citep{Li2019}. There is also evidence that concepts from chaos theory, such as the existence of strange attractors, can explain the dynamics of fully turbulent flows. The main challenge, however, remains the development of DA and ML methods that can deal with multi-scale nature and high dimensionality inherent to fully developed three-dimensional turbulence at high Reynolds numbers.

%
%
%




\backsection[Funding]{We gratefully acknowledge the discussions with Prof. Luca Biferale, Dr. Massimo Cencini and Dr. Yi Li. This work was supported by the National Natural Science Foundation of China (Grant Nos. 12225204 and 12002147), the Department of Science and Technology of Guangdong Province (Grant Nos. 2023B1212060001 and 2020B1212030001) and the Shenzhen Science and Technology Program (grant no. KQTD20180411143441009). We
acknowledge support from the Centre for Computational Science and Engineering at SUSTech.}

\backsection[Declaration of interests]{The authors report no conflict of interest.}


\backsection[Author ORCIDs]{V. Gupta, https://orcid.org/0000-0003-3990-9505; Y. Chen, https://orcid.org/0000-0003-4198-2939; M. Wan, https://orcid.org/0000-0001-5891-9579.}


\appendix

\section{Method-specific parameters}\label{Apar}

The three methods used in this paper are quite different from each other and hence has a different set of method-specific parameters. Therefore, we list them separately here for all the results obtained in this paper. The codes are also attached in supplementary material for the interested readers who wish to reproduce the results or understand these methods in more detail.
The main method-specific parameters for 4D-Var are $T_{assim}$, which should be approximately $0.5\Lambda_{max}^{-1}$ for optimal performance, and number of iterations ($N_{iter}$). To save the computational time, we sometimes used a threshold convergence $C_{conv} = J^i/J^{i-1}$ as a stopping criterion instead of $N_{iter}$. Another parameter used in 4D-Var is a regularisation factor $N_{reg}$, which is a Tikhonov regularisation used for obtaining the inverse of the Hessian. This means that the inverse of Hessian is obtained as $\left(\partial^2 J^i/\partial \mathbf{u}_0^{i2} + N_{reg}\mathcal{I} \right)^{-1}$, where $\mathcal{I}$ is the identity matrix.

The main method-specific parameters for EnKF are $T_{assim}$ and the number of ensemble members $N_{ens}$. The prediction accuracy of EnKF increases for longer $T_{assim}$ and higher $N_{ens}$, but that can significantly increase the computational time. We, however, keep $T_{assim}$ to be of the order $\Lambda_{max}^{-1}$, which is practical if the model has errors. Beyond some value, Increasing $N_{ens}$ has diminishing improvements. $N_{ens}$ is therefore kept in the order of $10^2-10^3$ to save the computational time. In order to use smaller $N_{ens}$, we also introduced regularisation in EnKF. Under this regularisation $\mathcal{O}_j$ becomes a diagonal matrix that assumes $N_{ens}$ to be infinite.

The main method-specific parameters for RC-RNN are the length of the training time $T_{train}$ and size of the reservoir network $D_r$. Longer $T_{train}$ and larger $D_r$ lead to better performance but needs to be limited to save computational time. We mention again that even in the training phase, the network only receives the sparse and noisy observations. The training is only performed once to fix the elements of $W_{out}$. The data during the training phase does not have any correlation with the system state during the prediction phase. In order to bring the reservoir to the system's state, the initialisation is performed using data in DAW. The length of the DAW, i.e. $T_{assim}$, is not an important parameter in RC-RNN~\citep{Pathak2018}. Other parameters in RC-RNN are the network hyper-parameters $\sigma$ and $\rho$, which determine the magnitude of elements in matrices $W_{in}$ and $A$, respectively. These hyper-parameters are optimised manually~\citep{Gupta2023}.

Figure \ref{ErK1}, for 4D-Var: ($T_{assim} = 5$, $N_{iter}=20$, and $N_{reg} = 10^{-2}$),
for EnKF: ($T_{assim} = 5$ and $N_{ens} = 400$),
and for ML: ($T_{train} = 8000$, $D_r  = 2000$, $\sigma = [0.01,0.005]$, and $\rho = 0.4$).
Figure \ref{SPvpt} (a), for 4D-Var: ($T_{assim} = 5$, $C_{conv} = 0.99$, and $N_{reg} = 10^{-2}$),
for EnKF: ($T_{assim} = 10$ and $N_{ens} = 100$ with regularisation),
and for ML: ($T_{train}=16000$, $D_r  = 2000$, $\sigma = [0.01,0.005]$, and $\rho = 0.4$).
Figure \ref{SPvpt} (b), for 4D-Var: ($T_{assim} = 2.4$; $C_{conv} = 0.99$, and $N_{reg} = 10^{-2}$),
for EnKF: ($T_{assim} = 10$ and $N_{ens} = 100$ with regularisation),
and for ML: ($T_{st} = 1$, $T_{train} = 16000$, $D_r  = 2000$, $\sigma = [0.01,0.005]$, and $\rho = 0.3$).
Figure \ref{SPvpt} (c), for 4D-Var: ($T_{assim} = 2$, $N_{iter}=200$, and $N_{reg} = 10^{-1}$),
for EnKF: ($T_{assim} = 3$ and $N_{ens} = 1200$),
and for ML: ($T_{train} = 800$, $D_r  = 2048$, $\sigma = [0.02,0.02]$, and $\rho = 0.2$).
Figures \ref{Err3} (a-c) and \ref{Err3FFT} (a-c), for 4D-Var: ($T_{assim} = 5$, $N_{iter}=20$, and $N_{reg} = 10^{-2}$),
and for EnKF: ($T_{assim}=5$ and $N_{ens} = 400$).
Figures \ref{Err3} (d-f) and \ref{Err3FFT} (d-f), for 4D-Var: ($T_{assim} = 2.4$, $N_{iter} = 20$, and $N_{reg} = 10^{-2}$),
and for EnKF: ($T_{assim} = 2.4$ and $N_{ens} = 400$).
Figures \ref{Err3} (g-i), \ref{Err3FFT} (g-i) and \ref{Ecgl}, for 4D-Var: ($T_{assim} = 2$, $N_{iter}= 100$, and $N_{reg} = 10^{-1}$),
and for EnKF: ($T_{assim}=2$ and $N_{ens} = 1200$).

\section{Fourier transform of the prediction errors for the CGL system}\label{AppB}

\begin{figure}
 \centerline{\includegraphics[width=0.9\textwidth, trim =5.3cm 0.4cm 6.7cm 1.55cm, clip]{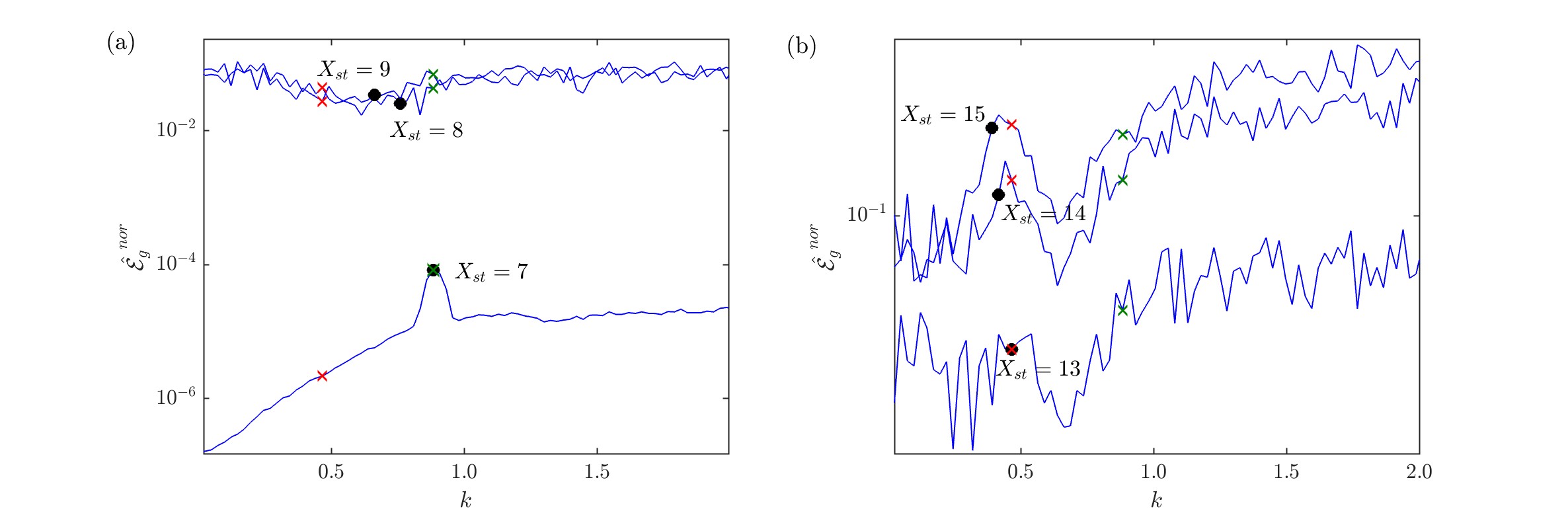}}
 \captionsetup{width=1.0\linewidth}
  \caption{$\hat{\mathcal{E}}_g^{nor}$ for the CGL system predictions by 4D-Var when spatial sparsity of measurements is (a) $X_{st} = 7,$ 8 and 9, and (b) $X_{st} = 13,$ 14 and 15. The green and red crosses correspond to $X_{st}$ at the two vertical dashed lines in figure~\ref{SPvpt}. The black circles correspond to $X_{st}$ for which $\hat{\mathcal{E}}_g^{nor}$ are plotted.}
\label{Ecgl}
\end{figure}

We further analyse the 4D-Var results for the CGL system, which are important for two reasons. First, they most clearly reveal the qualitative difference in the predictions between the different zones. Figure~\ref{Ecgl} shows $\hat{\mathcal{E}}_g^{nor}$ for several values of $X_{st}$ across the three zones. Panel (a) shows a sudden shift in the prediction errors between $X_{st} = 7$ and 8, i.e. transition from the good to reasonable-predictions zone. Panel (b) shows a sudden shift in the prediction errors between $X_{st}=13$ and 14, i.e. transition from the reasonable to bad-predictions zone. Second, these results illustrate the differences between 4D-Var and EnKF. In 4D-Var, we solve a nonlinear optimisation problem to minimise the cost function defined in the entire DAW. In EnKF, we consider a single measurement at each time and apply linear corrections sequentially. The large difference between 4D-Var and EnKF results for the CGL system in the middle rows of figures~\ref{Err3} and~\ref{Err3FFT} indicates that both methods can converge to different solutions. Interestingly, as seen in figure~\ref{Err3} (h), even though the error from 4D-Var is initially higher than that from EnKF, the growth of 4D-Var error is much slower. We caution that these results do not indicate the superiority of either method over the other.

\bibliographystyle{jfm}
\bibliography{DA}


\end{document}